\documentclass{ws-ijmpd}
\usepackage{graphicx,amssymb,aas_macros}

\usepackage[super,compress]{cite}

\usepackage[dvips,breaklinks]{hyperref}
\hypersetup{colorlinks,urlcolor=black,citecolor=black,linkcolor=black,filecolor=black}
\usepackage{breakurl}

\begin{document}

\newcommand{\beq}{\begin{equation}}
\newcommand{\eeq}{\end{equation}}  
\newcommand{\beqn}{\begin{eqnarray}}
\newcommand{\eeqn}{\end{eqnarray}} 
\newcommand{\gappr}{\stackrel{>}{\scriptstyle \sim}}
\newcommand{\lappr}{\stackrel{<}{\scriptstyle \sim}}
\newcommand{\degree}{\ensuremath{^\circ}}

\newcommand {\Msun} {{M$_{\odot}$ }}
\newcommand {\ergs} {erg~s$^{-1}$}
\newcommand {\ergcms} {erg\,cm$^{-2}$s$^{-1}~$}
\newcommand {\s} {s$^{-1}$}
\newcommand {\sax} {{\it Beppo}SAX }
\newcommand {\gpeak} {$\gamma_{\rm peak}$ }
\newcommand {\hess} {H.E.S.S.}
\newcommand {\gradi} {^{\circ} }
\newcommand {\nupeak} {$\nu_{peak}$ }
\newcommand {\es} {H\,1426+428 }
\newcommand {\C}{Cherenkov }
\newcommand {\gam} {$\gamma$}
\newcommand {\m} {$\mu {\rm m}$}
\newcommand {\rchisq} {$\chi_{\rm red}^{2}$}
\newcommand {\chisq} {$\chi^{2}$}
\newcommand {\nw} {nW m$^{-2}$ sr$^{-1}$}
\newcommand {\ggee} {$\gamma-\gamma \rightarrow e^+ e^-$}

\markboth{L. Costamante}{Blazars and EBL}

\catchline{}{}{}{}{}

\title{Gamma-rays from Blazars and the Extragalactic Background Light}

\author{Luigi Costamante}
\address{Department of Physics, University of Perugia,  I-06123 Perugia, Italy  \\
luigic2011@gmail.com}

\maketitle


\begin{history}
\received{17 July 2013}
\accepted{23 July 2013}
\end{history}

\begin{abstract}
Recent observations of blazars at high energy (HE, 0.1--100 GeV) and very high energy 
(VHE, $>$0.1 TeV)  have provided important constraints on the intensity and 
spectrum of the diffuse Extragalactic Background Light (EBL), shedding light on its main origin.
Several issues remain open, however, in particular in the mid- and far-infrared bands
and in the blazar emission at multi-TeV energies. 
This review  summarizes the observational and theoretical progress in the study of
the EBL with gamma-rays  and the most promising future improvements,
which are mainly expected from spectra in the multi-TeV range.
\end{abstract}

\ccode{PACS Nos.: 98.54.Cm,98.62.Ai,98.58.Jg,98.58.Ca,98.62.En,95.85.Hp,95.85.Pw,98.70.Vc}

\section{Introduction}	
Above an energy from few tens to few hundreds GeV, 
the Universe ceases to be  transparent to gamma-rays.
The intergalactic space is filled with the light 
produced by all the sources in the Universe throughout cosmic history,  
which in the band 0.1-1000 \m\ is called  the diffuse Extragalactic Background Light (EBL\cite{hauser}).
In this radiation field the photon-photon collision and pair production 
process becomes highly probable,
converting the original gamma-ray photons into pairs and lower energy photons.
While this represents a problem for the study of extragalactic gamma-ray sources, 
it also provides a method to probe
the EBL\cite{nikishov,jelley,gould,stecker92}.

\begin{figure}[t]
\includegraphics[width=0.5\textwidth]{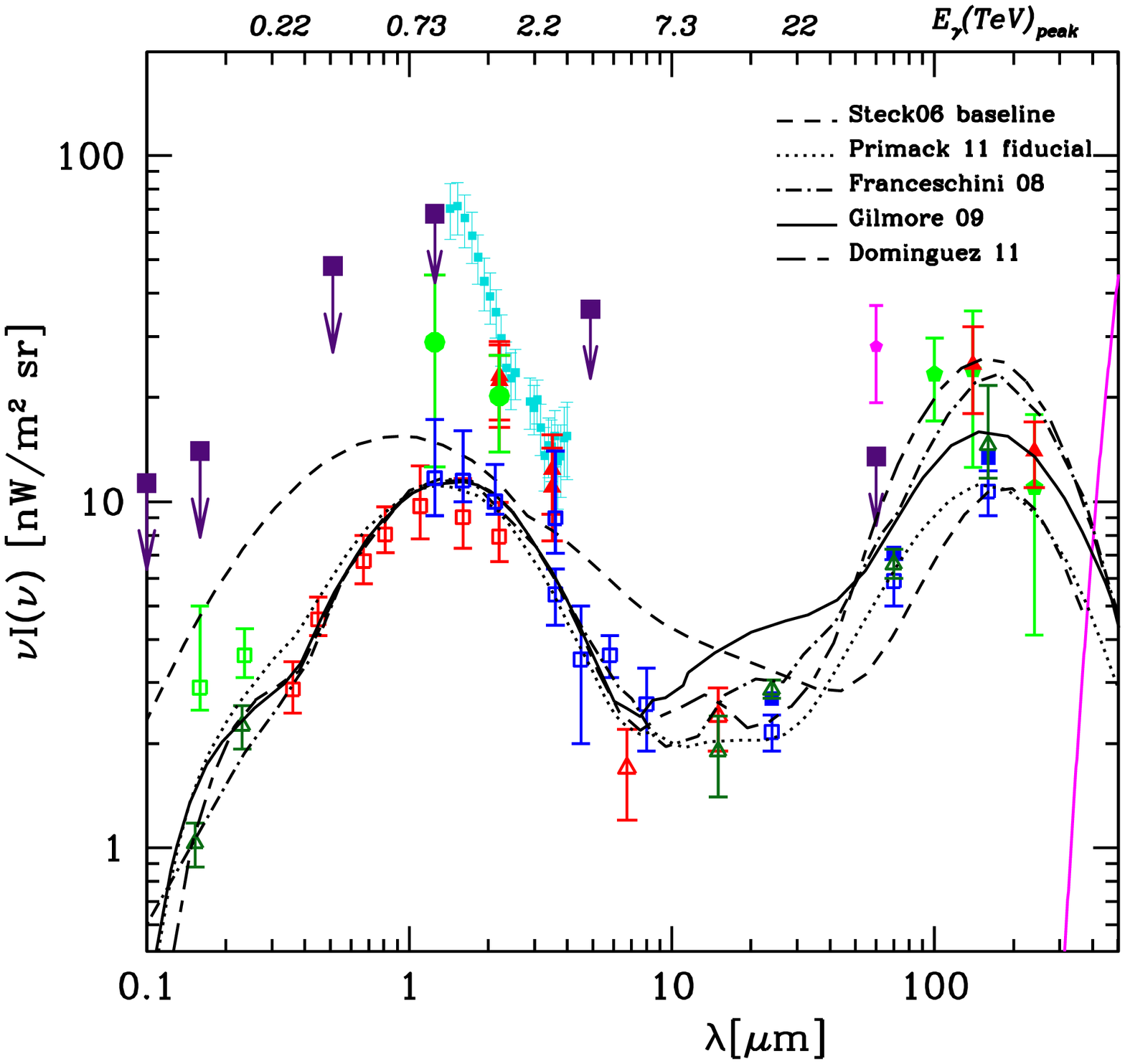}
\includegraphics[width=0.45\textwidth]{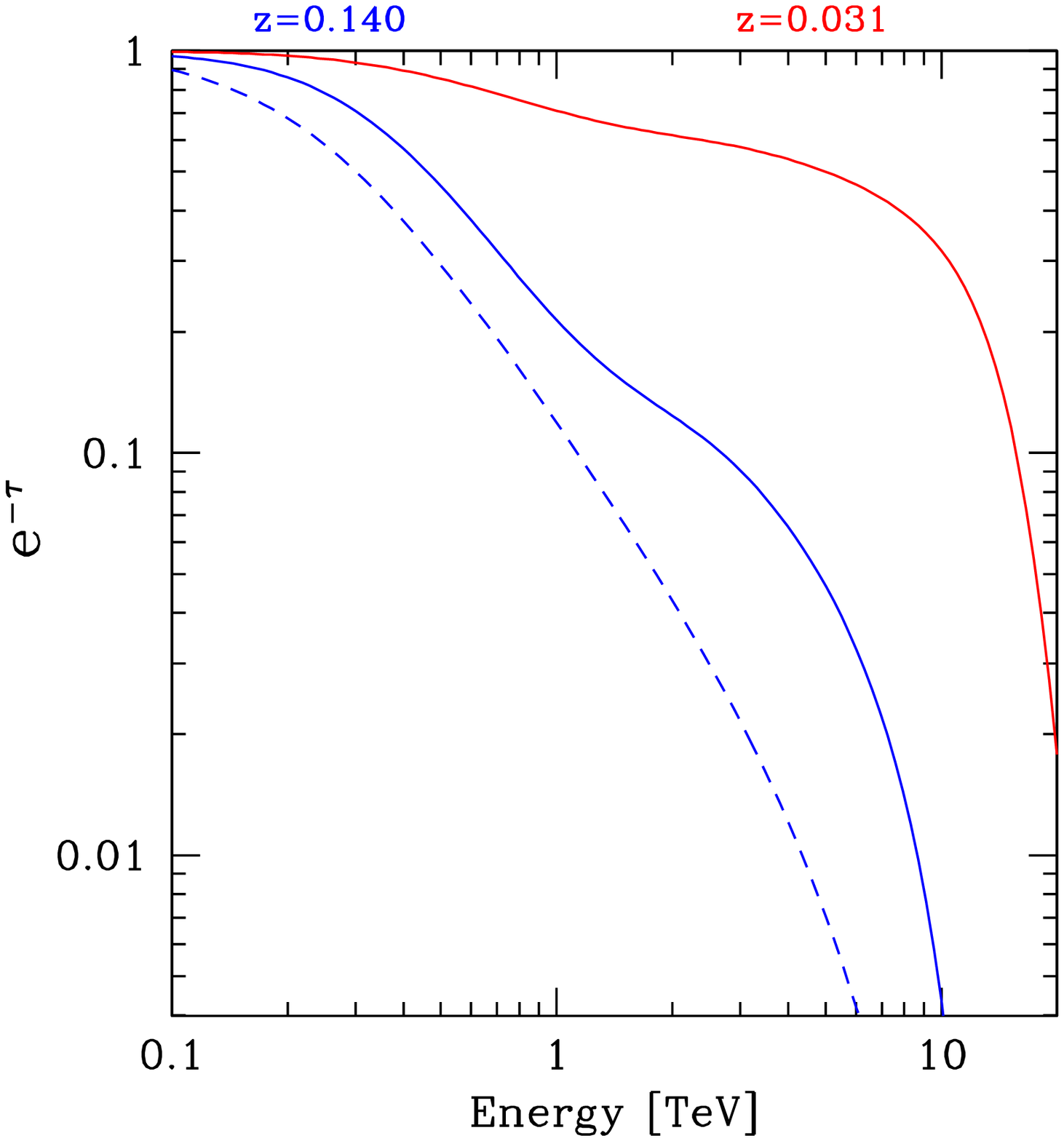}
\vspace*{-0.3cm}
\caption{Left: SED of the EBL. EBL data are from a review compilation\cite{hauser,nature_ebl},
unless otherwise stated.
Open symbols correspond to lower limits from 
galaxy counts\cite{gardner2000,xu_galex,madaupozzetti,keenan10,fazio04,dole06,levenson08}
while filled symbols correspond to direct estimates.
The highest initial estimates were from the NIR Spectrometer\cite{matsumoto} on IRTS,
between 1.5 and 4\m\ (cyan symbols) and from  COBE/DIRBE\cite{finkbeiner} 
at 60 \m\ (magenta symbol).
The curves show a sample of different recent EBL models, 
as labelled\cite{stecker06,franceschini08,gilmore09,primack11fid,dominguez11}.
On the upper axis it is plotted the TeV energy corresponding to the peak of 
the $\gamma-\gamma$  cross section.
Right: attenuation curves $e^{-\tau}$ for sources at two different redshifts (solid lines)
with the same EBL model\cite{dominguez11}, and at the same redshift (z=0.140)
for two different EBL models (dashed line for Stecker06\cite{stecker06}, 
solid lines for Dominguez11\cite{dominguez11}).  The attenuation curves can 
represent visually the shape of an observed VHE spectrum,  
if the intrisnic spectrum were a power-law parallel to the upper axis.
Calculations performed assuming $H_0=70\;\rm km/s/Mpc$, $\Omega_m=0.3$, $\Omega_{\Lambda}=0.7$.
}
\label{eblsed}
\end{figure}

The intensity and spectral shape of the EBL is a key issue in astrophysics,
providing unique information about the epochs of formation 
and the history of evolution of galaxies\cite{primack,dwek2013}.
The EBL spectral energy distribution (SED)  is expected to be dominated by two main humps\cite{hauser},
produced by the direct starlight from galaxies in the optical (Opt) to near- and mid-infrared (NIR-MIR) 
wavelengths,  and by the light reprocessed by dust in the mid- to far-infrared (MIR-FIR) 
wavelengths (Fig. \ref{eblsed}).
The integrated light from resolved galaxies down to the faintest end of the luminosity functions  
provides a strict lower limit\cite{gardner2000,xu_galex,madaupozzetti,keenan10,fazio04,dole06}.  
Direct estimates from the night-sky background 
show instead a higher isotropic emission, especially between 1.5 and 4 \m, 
which is often too bright to be accounted for by the integrated light of missed faint galaxies 
(see e.g. Ref. \citen{matsumoto,hauser}). This raised the issue of possible additional contributions 
from other sources, in particular by Population III (zero metallicity) 
stars\cite{santos,salvaterra,magliocchetti,kashlinsky04,cooray04} 
during intense star formation at $z\sim10$.  
However, the excess might not be extragalactic:  
direct measurements are affected by large systematic uncertainties
due to the difficulties in the accurate subtraction of the bright foregrounds\cite{hauser}, 
in particular zodiacal light\cite{dwek_pop3}.
In addition, the energetic requirements of the Population III hypothesis appear 
too high to be compatible with other properties of the present Universe 
like the average metallicity, soft X-ray background or the ionizing photon budget
from WMAP results\cite{madausilk}. 
This conclusion is confirmed
also by the comparison with scenarios of structure formation\cite{dwek_pop3}.

Gamma-rays provide a completely independent way to measure the EBL,
and blazars are the most luminous  persistent extragalactic 
gamma-ray sources in the Universe.

Blazars are radio-loud AGN with the relativistic jet pointed close 
to the line of sight\cite{urrypadovani}.
They are divided in flat-spectrum radio quasars (FSRQ) and BL Lacertae objects (BL Lacs)
according to the presence or absence of strong broad emission lines 
in their optical spectrum, respectively.
Their SED is characterized by two broad  humps, at lower and higher energies,
most commonly explained as synchrotron and inverse Compton (IC) emission (respectively) 
from a population of relativistic electrons in the bulk motion of the relativistic jet.
The seed photons for the IC process can be 
the synchrotron photons produced by the same electrons
(synchrotron self-Compton models, SSC\cite{gg85,bloom96}) 
or produced externally to the jet,
either directly by the accretion disk\cite{dermer92}
or reprocessed  by clouds in the Broad Line Region (external Compton model, 
EC\cite{sikoraec94}), or produced by the dust torus 
in the AGN (external Compton on IR dust emission, EC(IR)\cite{sik_mevbl02}).
Seed photons can come also from other parts of the jet\cite{ggspinelayer,markos03}.
The blazars' SEDs display a large, continuous range of synchrotron peak frequencies,
and can be broadly divided into low and high-energy peaked objects, according if the 
synchrotron \nupeak\ is
in the IR/optical or in the UV/X-ray bands\cite{giommipadovani94,1LAC},
and correspondingly the IC peak falls in the HE or VHE bands\cite{1LAC,2LAC}.

Blazars represent an excellent class of extragalactic $\gamma$-ray beamers, 
being luminous sources from GeV to multi-TeV energies,
and well distributed over a wide range of distances ($z\sim0-4$). 
However, they are not standard candles: their emission is highly variable 
in unpredictable way\cite{variab_urry,variab_lat10,BIGFLARE}, 
their SED span a wide range of properties\cite{giommipadovani94,fossati98,ggsequence2}.
and the jet composition and emission mechanisms are not yet fully known\cite{felixbook04,anita13}. 

Gamma-gamma absorption establishes a strict one-to-one relation between 
the EBL and the VHE intrinsic spectrum, for any given redshift. 
The study of the EBL through gamma-rays constitutes 
a classic problem of one equation with two variables\cite{felixicrc} (even three if redshift is unknown).
Gamma-ray data have thus been used to 
1) constrain the EBL by assuming an intrinsic spectrum and redshift;
2) constrain blazar properties by assuming an EBL and redshift;
3) constrain the redshift by assuming an EBL and intrinsic blazar spectrum.

The advent of a new generation of Cherenkov telescopes 
(H.E.S.S.\cite{hess}, MAGIC\cite{magic}, VERITAS\cite{veritas}) 
and the Fermi satellite\cite{fermi}
have increased dramatically the number of sources and quality of spectra at both
HE and VHE bands\footnote{For an up-to-date list see the online TeV Source Catalog 
(TeVCat) at http://tevcat.uchicago.edu/}, 
narrowing down the parameters space and  
allowing a quantum leap in our understanding of this issue.

\section{EBL modifications to the gamma-ray spectra}
\label{diagn}
The intrinsic gamma-ray spectrum $F_{int}(E)$ emitted by a source
is modified by EBL absorption 
because of the energy dependence of the  optical depth $\tau_{\gamma\gamma}(E,z)$.
If the intergalactic magnetic field is large enough to spread 
the secondary emission from the produced pairs over more than 3-4\degree  
(the typical field of view used in Cherenkov telescopes for background subtraction),
the observed spectrum of an extragalactic source is simply given by 
$F_{obs}(E)=F_{int}(E) exp[-\tau_{\gamma\gamma}(E,z)]$.  

The typical attenuation curve $e^{-\tau}$ for different shapes of the EBL spectrum and redshifts
is shown in Fig. \ref{eblsed}.
Several features can be noted (see also Ref. \citen{preston}):

1) because $\tau_{\gamma\gamma}$ mostly increases with energy for all expected EBL shapes,
EBL absoprtion makes the observed spectrum steeper than the intrinsic one 
(namely, if $F(E) \propto E^{-\Gamma}$,  the  photon index 
$\Gamma_{obs}\geq\Gamma_{int}$).
This steepening increases with redshift and EBL intensity. 

2)  The attenuation curve flattens    
between $\sim$1-2 and 8-9 TeV,  
for all recent EBL calculations\cite{franceschini08,finke10,dominguez11,gilmore12}.  
This is caused by the slope of the EBL spectrum between  2 and 10 \m, 
i.e. in the declining part of the ``valley"  between the stellar and dust emission humps. 
For a power-law spectrum of the EBL number density $n(\epsilon)\propto \epsilon^{-\beta}$
(i.e.  $\nu I_{\nu}\propto \lambda^{\beta-2}$ in the SED),
the optical depth becomes proportional to $\tau_{\gamma\gamma}\propto E^{\beta-1}$. 
Where $\beta\approx 1$, the optical depth becomes independent of energy\cite{guy2000,felixicrc}.
In such case there is no steepening or modification of the spectrum: 
the observed (absorbed) spectrum reproduces the intrinsic shape, simply attenuated by a constant factor.
Therefore an observed gamma-ray spectrum can appear harder in the 2-8 TeV range 
than in the 0.2-1 TeV band, even if its intrinsic spectral index is  constant over the whole energy range.
This flattening feature in the attenuation curve disappears if the EBL  slope is $\beta>1$, 
as for example in the model by Stecker 2006\cite{stecker06}, for which a continuous steepening 
is expected (Fig. \ref{eblsed}).
This flattening feature is not a consequence of the large width of the $\gamma$-$\gamma$
cross section, but only of the slope of the spectrum of target photons.

3) As a result,  NO cutoff is expected from EBL absorption between 0.2 and $\sim$8 TeV.
The only two cutoffs caused by EBL attenuation occurs in the transition between HE and VHE bands, 
around 100-200 GeV in the local universe 
(and decreasing towards higher redshift, see Fig. \ref{eblsed}), 
and above $\sim$8 TeV.   They are both caused by the EBL flux rising in the SED, 
towards the stellar and dust emission peaks respectively. 

4) The steepening of the gamma-ray spectrum  is essentially determined by the contrast in optical depth 
between the two ends of a detected VHE band.
Any spectral steepening therefore can always be canceled  
by assuming a higher EBL flux at shorter wavelenghts, 
so to make the EBL spectrum closer to the $\lambda^{-1}$ behavior and thus equalize the optical depths
\cite{nature_ebl,preston}.

\begin{figure}[t]
\raisebox{0.9cm}{\includegraphics[width=0.46\textwidth]{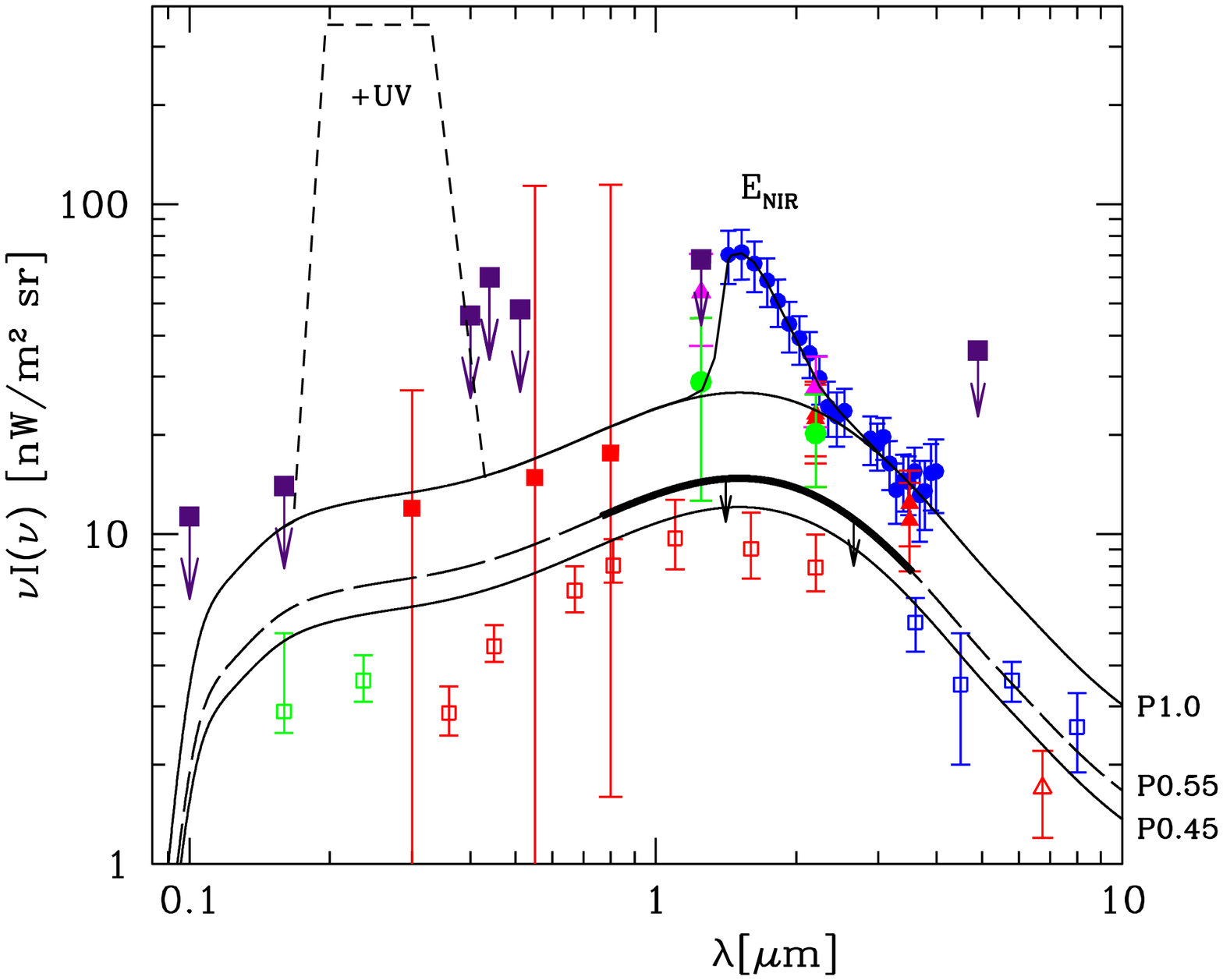}}
\includegraphics[width=0.53\textwidth]{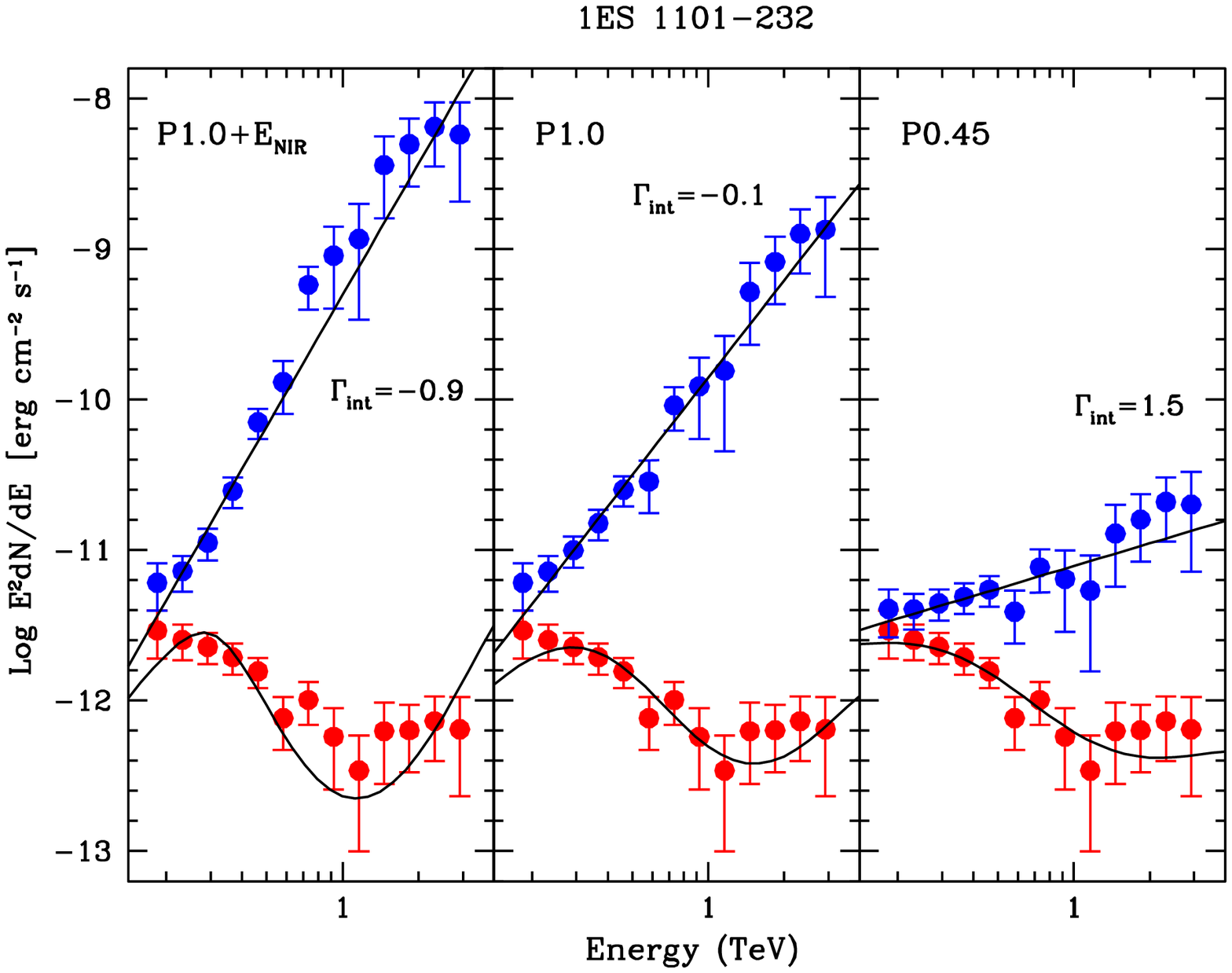}
\vspace*{-1.5cm}
\caption{Left: SED of the EBL with test EBL shapes and upper limits curves
derived from the assumption of an intrinsic spectrum $\Gamma_{int}\geq1.5$ (black marked region).
Right: the H.E.S.S. spectrum of 1ES\,1101-232, corrected for absorption 
with three different EBL shapes, as labeled.
It shows how reducing the EBL intensity softens the required intrinsic spectrum.
Lower (red) points: observed data. Upper (blue) points: 
absorption-corrected data. The lines show the best fit power-laws to the 
reconstructed spectrum, and the corresponding shape after absorption.
Figures from Ref. \citen{nature_ebl}.
}
\label{nature}
\end{figure}

\section{Constraints from the VHE spectra of blazars}
To put constraints on the EBL, it is necessary to make assumptions 
on the blazar intrinsic spectrum. 
The different EBL limits obtained so far can be divided in three broad classes,
according to the different methods used to infer the intrinsic blazar spectrum.
The main assumptions are based on: \\
1) the hardness of the VHE spectrum, and in general the absence of exponential up-turns or
pile-up features; \\
2) the extrapolation of the Fermi-LAT spectrum to the VHE band; \\
3) the SSC/EC modeling of the full SED.

\subsection{Hardness of spectrum}
\label{irb}
One of the most general and robust approaches to EBL limits 
is to require the reconstructed  gamma-ray spectrum to simply stay within 
the known range of properties of the source\cite{costamante04}.
Since the spectral steepening increases with EBL intensity, 
upper limits can be derived by adopting an hardness limit  
on the reconstructed gamma-ray spectrum, comprising the absence of sharp pile-up features.

Blazars show a wide range of spectra, but are typically characterized by 
$\Gamma\gtrsim1.5$ at frequencies corresponding to the highest particle energies\cite{nature_ebl}.
These values can be easily produced with standard shock acceleration 
and cooling mechanisms. Other scenarios with much harder particle spectra are theoretically possible,
and have been studied in the context of blazars
even producing pile-up or line-like emission features 
(see e.g. Ref. \citen{felixicrc,lefa11} and references therein).
However, they have never been significantly detected in blazars so far,  
neither directly or by synchrotron emission,
requiring a cosmic conspiracy to avoid conflicts with multi-wavelength SED data.
The photon index $\Gamma$=1.5 can therefore be used as reference value, representing so far
the borderline between reality and speculation (see discussion in Ref. \citen{preston}). 

The EBL upper limits from spectral hardness are in general not affected by cascades 
in the intergalactic medium. The electromagnetic cascade initiated by the primary gamma-rays 
transfer the energy output from the absorbed to the transparent part of a gamma-ray spectrum 
(i.e. from high to low energies at VHE). 
Even assuming a null intergalactic magnetic field (IGMF), therefore, the addition 
of the full cascade contribution further steepens the observed VHE spectrum 
(see Fig. 2 in Ref. \citen{vovk0229}),  making the EBL limits even more constraining.

Similar considerations are valid when there is additional gamma-gamma absorption 
in the source, if the target photon field is sufficiently broad-banded.
Internal absorption on a broad-band target field reinforces the EBL attenuation curve,
and the resulting additional steepening would make the EBL limits even more stringent.
The exception is if the target field is narrow-banded, 
for example as a Planckian distribution\cite{akc08}.
In such case the optical depth can also decrease with energy, in some spectral band.
Internal gamma-gamma absorption can thus make the emerging gamma-ray spectrum 
even {\it harder} than the emitted one. This can be an effective way to produce 
very hard spectra without assuming hard particle distributions, at the expense of 
source luminosity\cite{akc08}. 
The absence of strong fluxes in the Fermi-LAT band (which should be transparent in this picture) 
does not lend support to this scenario for the hard TeV BLLacs, but for other BL Lacs 
with high Fermi-LAT fluxes it could provide an elegant explanation for observed mis-matches 
between HE and VHE spectra (see e.g. Ref. \citen{olga}).

\subsubsection{Optical-NIR limits.}
At wavelengths up to 1-2 \m, no meaningful limit (i.e. below the direct estimates) to the EBL 
could be derived until 2005, 
because of the general softness or low redshift of the detected sources. 
A measured VHE spectrum could be the result either of a hard intrinsic spectrum 
attenuated by a high-density EBL, or of a soft intrinsic spectrum less attenuated by a low-density EBL.  
In both cases the required intrinsic slope up to 1-2 TeV was within the typical range expected in blazars.
A low EBL density did however seem preferable, yielding a more typical SEDs 
for the detected objects\cite{costamante04}. 
The first H.E.S.S. spectrum on PKS\,2155-304 provided another possible hint,
requiring a sharp upturn and hard rise in case of the highest EBL estimates at 1-5 \m,
but only if the EBL in other wavebands were kept as low as galaxy counts\cite{dwek2155}.
These anomalous features disappear considering a slightly higher overall 
EBL level (especially in the UV part), which would cancel the inflection break and excess hardness.
Neglecting the EGRET spectrum, taken more than 10 years before and of low significance,
the resulting SED could still be typical for a blazar, with an IC peak at few TeV\cite{hess2155}.
The PKS\,2155-304 data therefore could not rule out an extragalactic origin for the 1-5 \m\ excess.

The breakthrough was achieved in 2005\cite{nature_ebl}, with the H.E.S.S. detections of 1ES\,1101-232 ($z$=0.186) 
and H\,2356-309 ($z$=0.165). 
The observed $\gamma$-ray spectra (measured between 0.2 and 1-3 TeV) were much harder 
than expected for their redshift,  implying extremely hard intrinsic spectra ($\Gamma_{int}\lesssim 0$) 
in case of high EBL densities (Fig. \ref{nature}).
To accomodate  $\Gamma_{int}\geq1.5$, the  upper limit on the EBL becomes
$\lesssim(14\pm4)$\,\nw\  around 1--2 \m, which is 
very close to the lower limits given by the integrated light of galaxies\cite{nature_ebl},
and even closer to the most recent NIR estimates\cite{keenan10}.
This means that more than 2/3 of the EBL in the Opt-NIR band is resolved into single sources.

Remarkably, the leverage given by 
larger-than-before redshift  and the good statistics of the data 
make this result robust against different EBL shapes (as long as it peaks arouns 1-2 \m),
systematic errors or small changes in the blazars assumption\cite{nature_ebl}.  
In the case of 1ES 1101-232, even assuming $\Gamma_{int}\sim 1.2$ affects the EBL limit 
by no more than 10\%.
This result has now been further corroborated by the VHE spectra of several 
other sources and datasets\cite{hess0347, veritas1218, magic3c279}.
It finally demonstrates that  1) the EBL up to 2-3 \m\ is dominated by the direct starlight 
from galaxies, excluding a strong contribution from other sources like Pop-III stars;
2) the intergalactic space is consequently more transparent to $\gamma$-rays than previously thought, 
thus enlarging remarkably the $\gamma$-ray horizon\cite{nature_ebl}.

\begin{figure}[t]
\raisebox{0.25cm}{\includegraphics[width=0.46\textwidth]{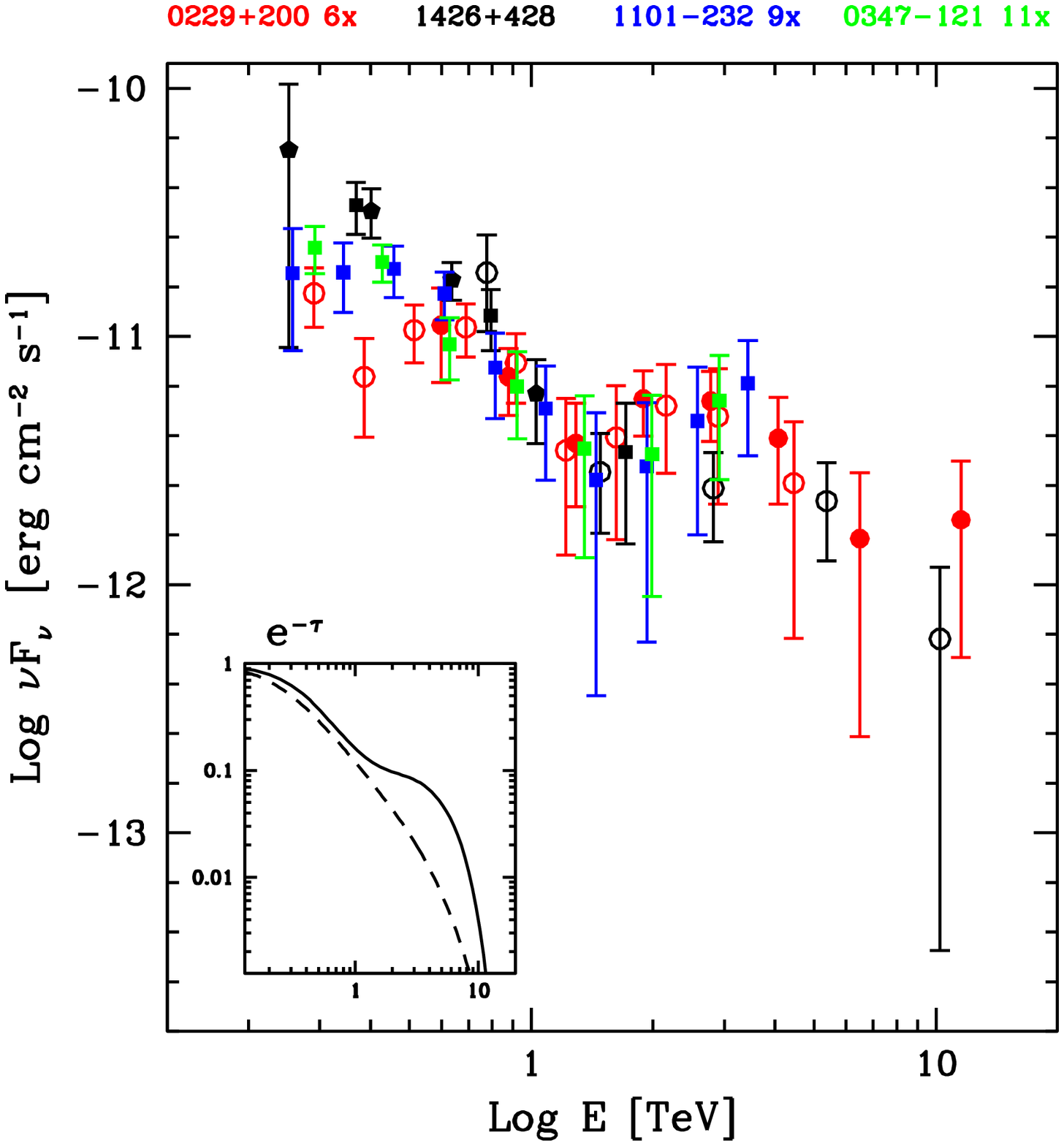}}
\includegraphics[width=0.48\textwidth]{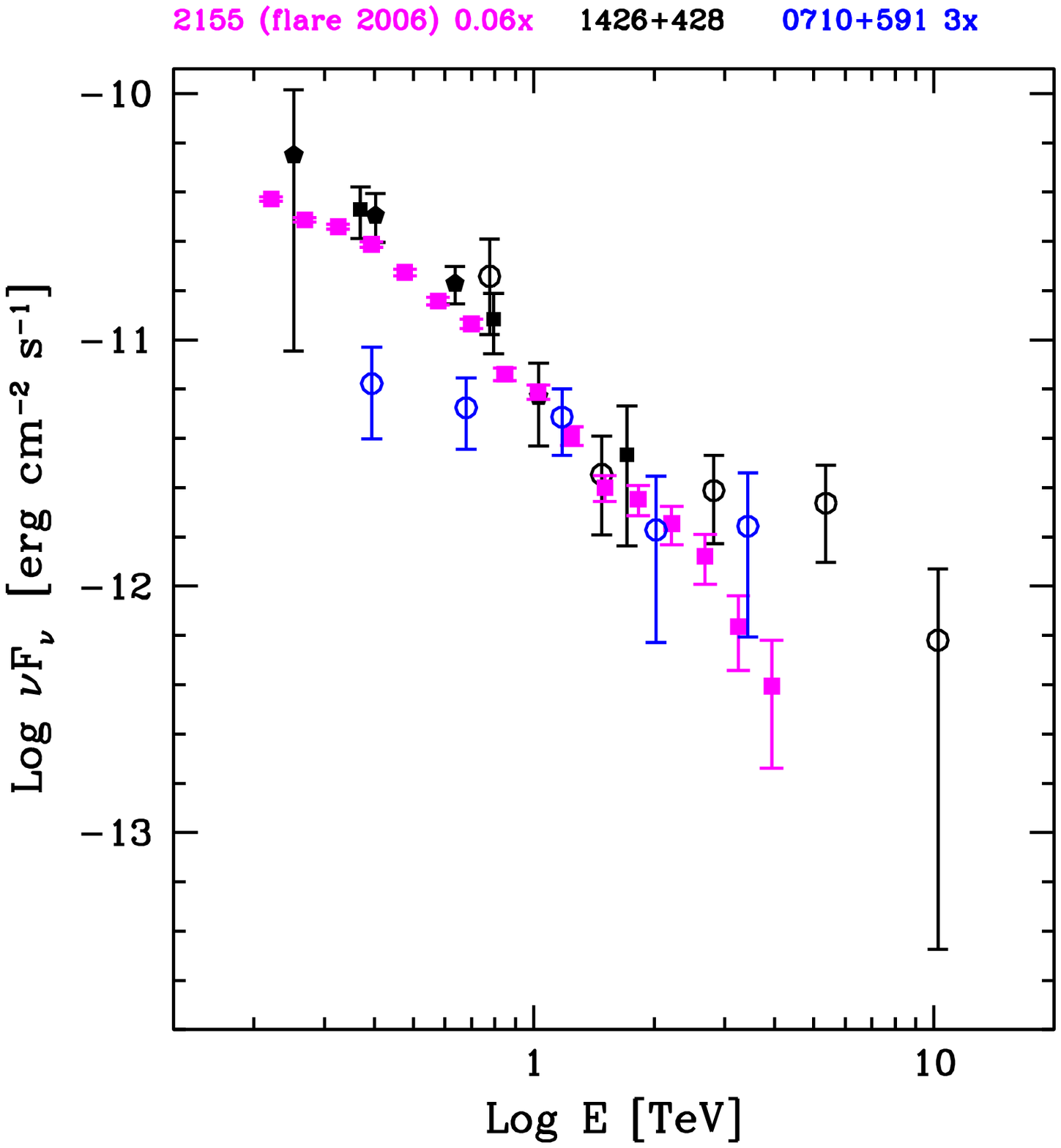}
\vspace*{-0.5cm}
\caption{Observed VHE spectra of all objects detected above 1-2 TeV, 
rescaled as labeled to match the flux of 1ES\,1426+428 around 0.8-1 TeV.
The data points above 1 TeV lie above the extrapolation from lower energies,
showing the flattening feature  expected from $\gamma-\gamma$ absorption with an EBL spectrum 
as given by standard galaxy evolution 
models\cite{franceschini08, primack11fid,dominguez11,gilmore09,gilmore12}
and not as predicted by the Stecker06\cite{stecker06} shape (see inset).
Left: data for 1ES\,1426+428\cite{hegra1426b,whipple1426,cat1426}, 
1ES\,0229+200\cite{hess0229,murase12}, 1ES\,1101-232\cite{nature_ebl} and 1ES\,0347-121\cite{hess0347}. 
Right: data for PKS\,2155-304\cite{chandranight} from the flare in 2006,
and RGB\,J0710+591\cite{veritas0710}.
Adapted and updated from Ref. \citen{cracovia}.
}
\label{flattening}
\end{figure}

\subsubsection{NIR-MIR limits.}
The first meaningful limits 
in this band could be derived with the HEGRA data on 1ES\,1426+428, which was 
the first object detected above 1 TeV beyond z=0.1.
The HEGRA spectrum\cite{hegra1426,hegra1426b} was significantly harder than measured
below 1 TeV\cite{whipple1426,cat1426}.
To avoid $\Gamma<1.5$ or hard upturns in the 1ES\,1426+428 TeV spectrum, 
the EBL cannot be as high as $\sim60$ \nw\ around 1.5 \m, or have
a slope much flatter than $\propto \lambda^{-1}$ between 2 and 10 \m\ \cite{hegra1426b},
as in the EBL calculation by Ref. \citen{malkanstecker01}.
Combining the data of all experiments available at the time (HEGRA, WHIPPLE, CAT),
the overall 0.2-TeV spectrum of 1ES\,1426+428  showed for the first time the typical 
flattening feature expected from a starlight-dominated EBL spectrum,
instead of the further steepening expected from a flatter EBL spectrum\cite{malkanstecker01,stecker06} 
(see Fig. \ref{flattening}).
The result however was affected by a limited statistics and the necessity 
to use three different experiments (HEGRA, WHIPPLE, CAT) to cover the full band from 0.2 to 10 TeV.

A stronger confirmation came with
the H.E.S.S. detection of 1ES\,0229+200\cite{hess0229} ($z=0.140$), 
whose spectrum was measured with good statistics up to $\sim10-15$ TeV. 
Taking advantage of the new EBL limits at 1-2 \m,
the  1ES\,0229+200 data 
constrains the EBL SED between 2 and 10 \m\  to be $\propto\lambda^{-\alpha}$
with $\alpha\gtrsim 1.1\pm0.25$ \cite{hess0229}, 
yielding an upper limit flux very close to the lower limits given by galaxy counts from Spitzer 
\cite{fazio04,dole06}.
This is possible because the Opt-NIR limit precludes the possibility of
reducing the contrast in optical depth between 1 and 10 TeV by 
considering a higher EBL at $\sim$1 \m\  (see Sect. \ref{diagn}).
The 1ES\,0229+200 constraint is also confirmed  by another analysis, 
based on the expected intensity of the flattening break in the VHE spectrum around 1-2 TeV 
produced by the EBL\cite{orr11}.

The 1ES\,0229+200 and 1ES\,1426+428 results are now being corroborated by an 
increasing number of objects,  when the VHE spectrum is measured above 1-2 TeV (Fig. \ref{flattening}).
Compared to the spectrum up to 1 TeV, the data points of all sources  
lie systematically above the extrapolation of the spectral slope at lower energies,
following the expected attenuation curve given by a pure starlight EBL.
Even when the intrinsic shape of the spectrum is not a power-law, 
and a curved spectrum can partly cancel the evidence as in the case of PKS\,2155-304,
the flattening feature is still visible when the statistics is large,
and is confirmed by the smoothness of the intrinsic blazar spectrum when reconstructed 
with standard EBL models\cite{chandranight}.

\begin{figure}[t]
\includegraphics[width=0.48\textwidth]{felix_icrc_f11.eps}
\includegraphics[width=0.48\textwidth]{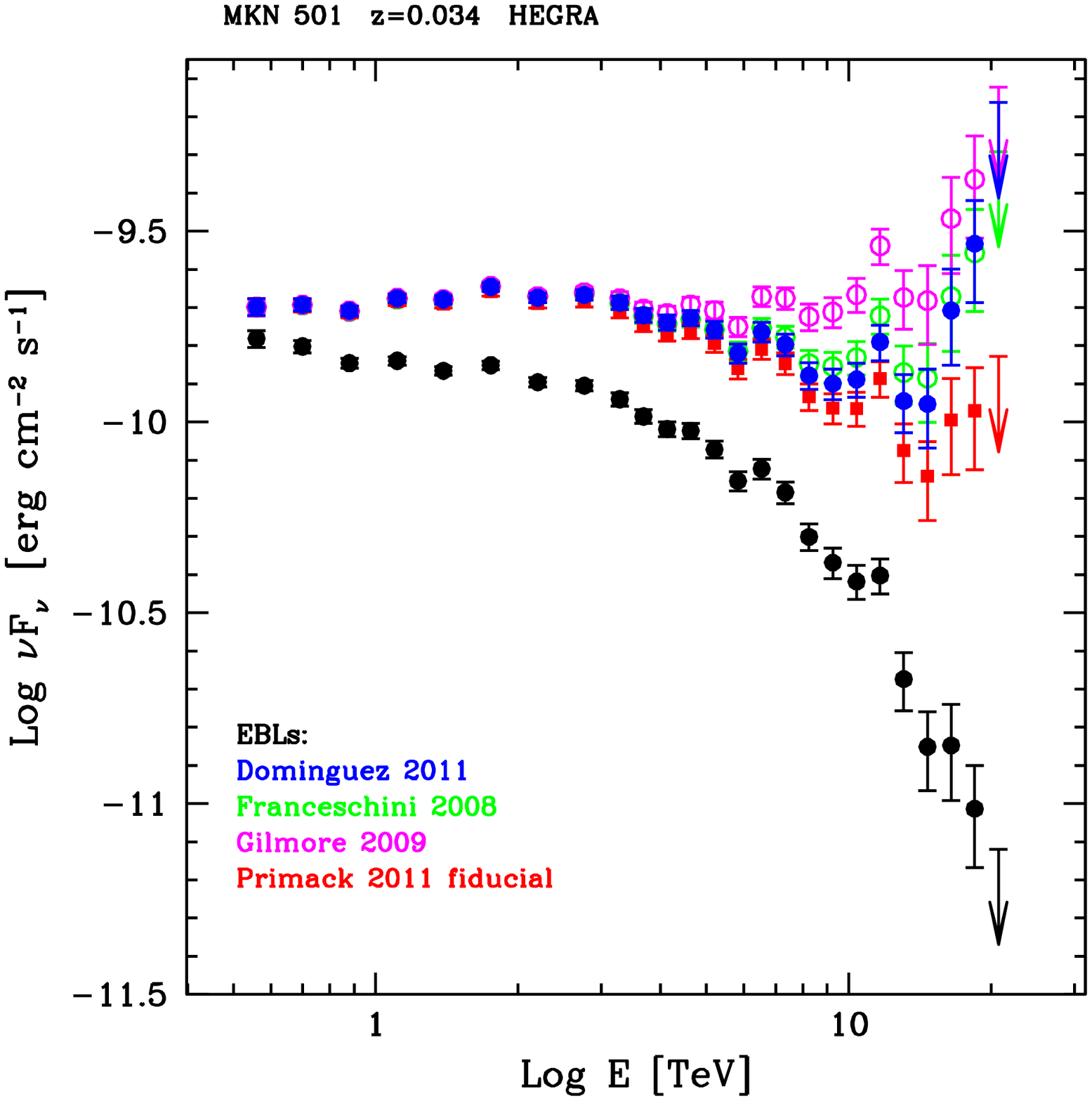}
\caption{``Ancient" and ``modern" versions of 
the ``IR background -- TeV gamma-ray" crisis\cite{ircrisis1}.
Left: models of pile-up spectra to explain the HEGRA data of Mkn\,501, 
if the highest 60 \m\ EBL flux\cite{finkbeiner} were right.  From Ref. \citen{felixicrc}.  
Right:  HEGRA data as observed\cite{hegra501rean} (black, lower points) and corrected for absorption
with  4 recent EBL models, as labeled\cite{franceschini08,gilmore09,dominguez11,primack11fid}.
Except for the ``Primack 2011 fiducial\cite{primack11fid}" model,
all seems to cause a sharp up-turn in the HEGRA spectrum. }
\label{ircrisis}
\end{figure}

\subsubsection{MIR limits and beyond}
At longer wavelengths, the limit at 10 \m\ around 3.1 \nw\  severely constrains the rising part 
of the EBL SED towards the far-infrared hump. This wavelengths are probed by TeV spectra above 10 TeV,
and the best measured blazar spectra above these energies are still represented by the HEGRA data on 
Mkn\,421\cite{hegra421} ($z$=0.031) and Mkn\,501\cite{hegra501spec,hegra501rean} ($z$=0.034).
The first EBL constraints were derived in relation to an initial high estimate\cite{finkbeiner}
of the EBL at 60 \m. To avoid a sharp pile-up at the end of the TeV spectrum
of the two Markarians, in particular Mkn\,501\cite{hegra501spec,felixicrc} (see Fig. \ref{ircrisis}), 
the EBL should be much less than the estimated $\sim30$ \nw, 
a problem known as  ``IR background -- TeV gamma-ray crisis\cite{ircrisis1}"
The crisis was resolved as a likely overestimate of the EBL.
With the new strong limits at 10 \m, however, there is less room for a compatible EBL.
In fact, the Mkn 501 data seem to have problems with the most recent EBL calculations, 
which cause again an up-turn or pile-up at the highest energies (see Fig. \ref{ircrisis}). 
This could turn out to be a modern version of the ``IR background -- TeV gamma-ray crisis",
but the information on the warm-dust emission and the  $\gamma$-ray statistic
are not sufficient to draw any sensible conclusion. 
More data above 10 TeV are needed, hopefully with the present 
or next generation of telescopes (e.g. CTA\cite{cta}).

\subsection{Combining constraints from multiple VHE spectra}
The EBL photon field, evolving with redshift, should be essentially common to all sources.
Each detected source provides a different combination of distance and gamma-ray energy band.
The limits from different sources  can therefore be combined 
to constrain the EBL over a wider range of wavelengths  and in a more stringent way 
than allowed by each single spectrum.

As shown in Sect. \ref{irb}, 0.1-1 TeV spectra constrain the EBL optical peak, 
1-10 TeV spectra constrain the EBL between 2 and 10 \m, and $>$10 TeV spectra 
probe the EBL beyond $\sim10$ \m.
Each new spectrum can take advantage of the previous limits, forming
a chain of constraints that starts with (and depends on) the assumption on the UV flux.
This is because the steepening of the gamma-ray spectrum can always be counteracted by
an increase in optical depth at the low-energy end of the detected gamma-ray band,
by means of a higher EBL flux at short wavelengths.
The most constraining sources are typically those with the hardest and most precise spectrum 
measured in each relevant band.

\begin{figure}[t]
\includegraphics[angle=-90,width=0.40\textwidth]{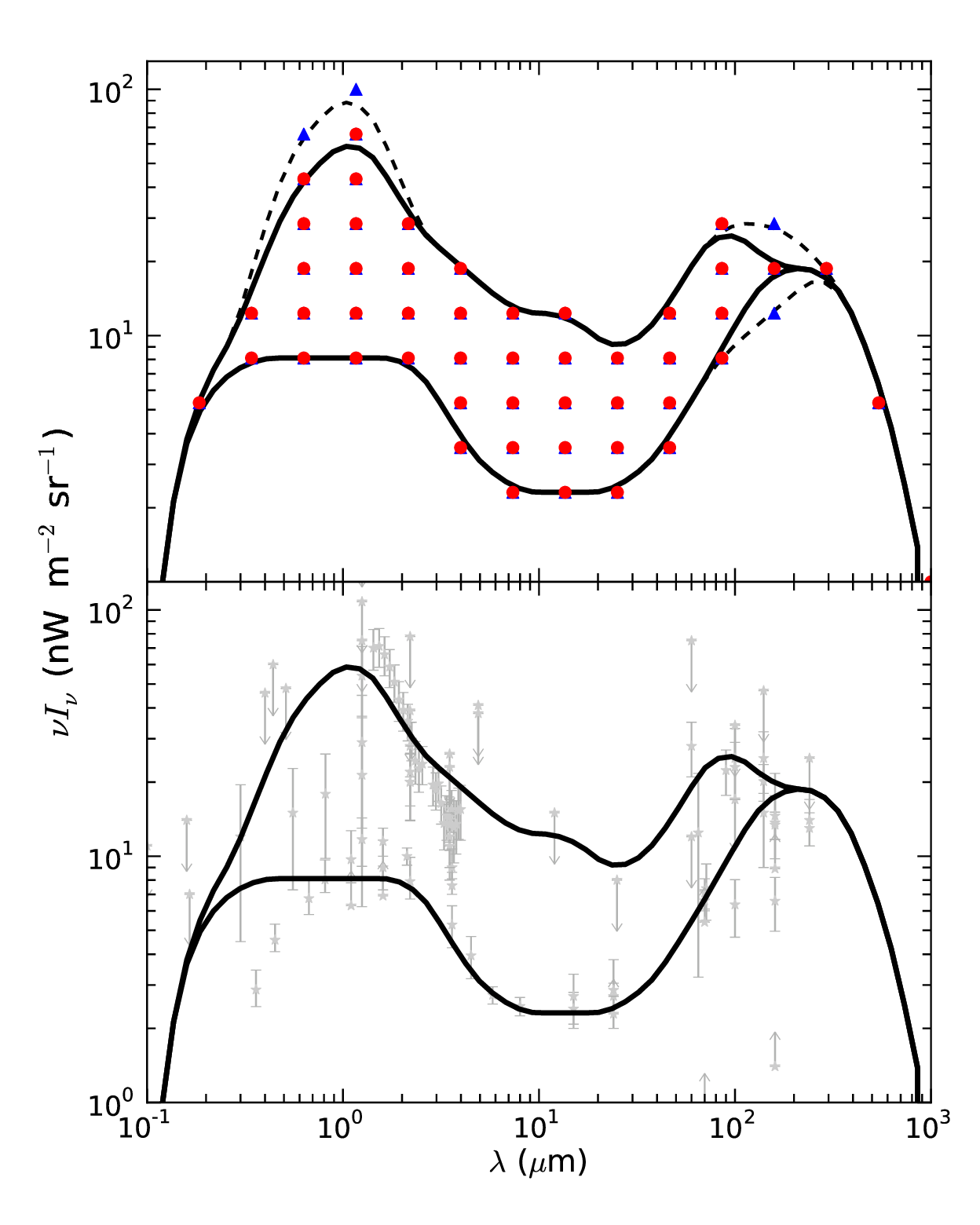}
\includegraphics[angle=-90,width=0.59\textwidth]{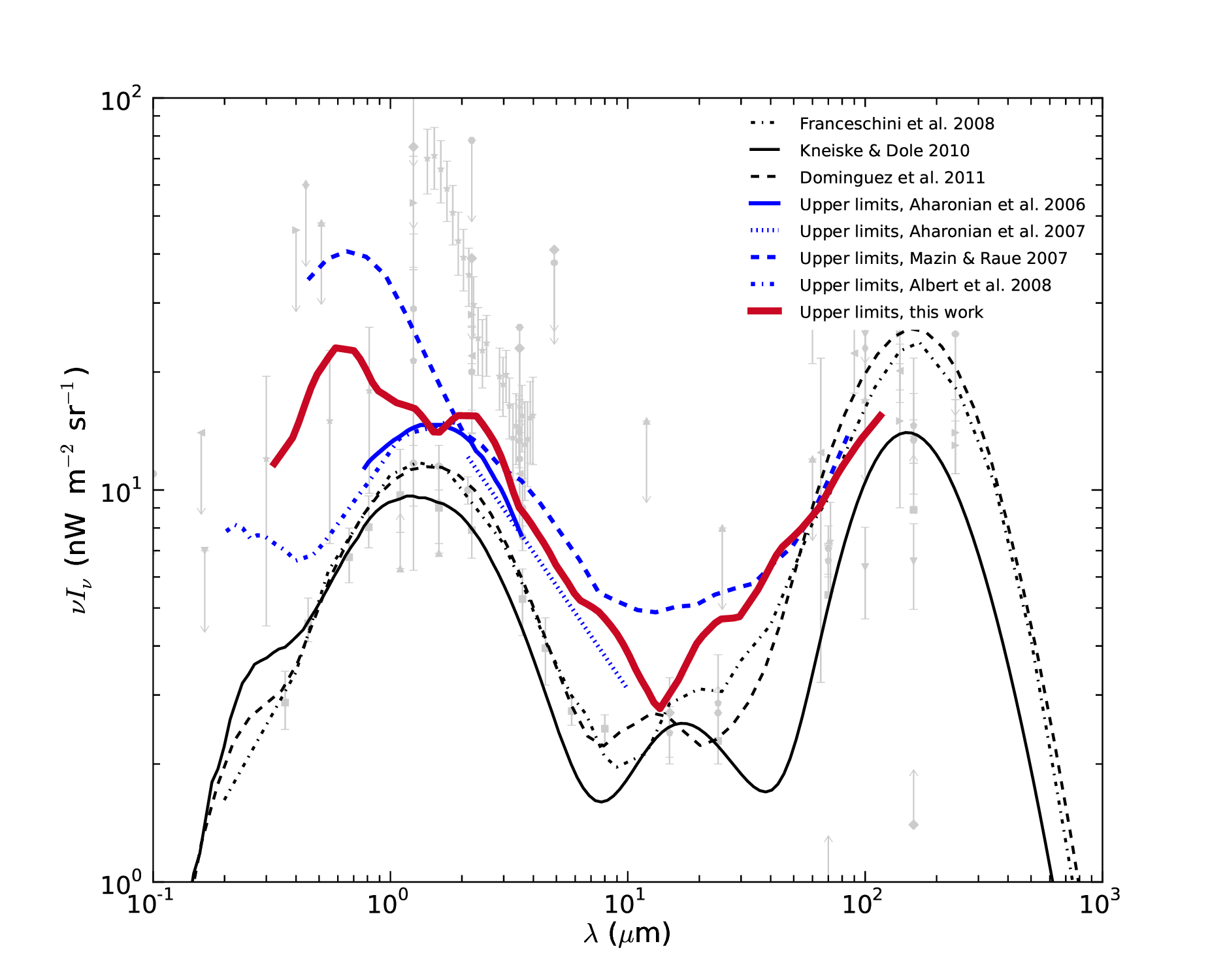}
\caption{Left: grid of the EBL energy flux vs wavelength space
used to construct the EBL shapes with splines for testing with VHE spectra (upper panel). 
Also shown are the minimum and maximum shapes tested 
(solid lines for Ref. \citen{meyer} and dashed lines for Ref. \citen{mazinraue}), 
together with the EBL limits and measurements (lower panel, light grey).  
Right: upper-limit curve derived as envelope of the allowed shapes in the left-panel grid, 
for $\Gamma_{int}\geq1.5$, together with previous limits and EBL models as labeled. 
All plots from Ref. \citen{meyer}. }
\label{meyer}
\end{figure}

This approach was developed by Ref. \citen{costamante04} and later expanded by Ref. \citen{dwek05},
with the goal of finding the highest viable EBL SED compatible with 
the assumption of $\Gamma_{int}\geq1.5$ for all blazars spectra.
As more objects became available, the analysis was updated by Ref. \citen{mazinraue}, 
but adopting a different goal and procedure. Instead of the highest viable EBL,
an upper limit curve is derived corresponding to the highest possible flux at each EBL wavelength.
This is achieved by scanning all possible smooth EBL shapes drawn as splines inside 
a selected grid of points, in the energy-density vs wavelength space (Fig. \ref{meyer}).
Then the envelope of all the highest allowed shapes is calculated. 
The grid of points sampling the EBL space is kept limited, in particular in the UV band,
otherwise the procedure would not converge (see Sect. \ref{diagn}).
The derived upper-limit curve depends on the choice of the grid points in the UV band,
i.e. on the assumption of a low UV EBL flux, as for the other analyses.

The resulting curve does not provide a viable, consistent EBL SED 
(in the sense that the envelope of the shapes from the $\Gamma_{int}\geq1.5$ limit  
can give $\Gamma_{int}<1.5$ when applied to single sources, e.g. for 1ES\,1101-232), 
but tells that, at any given EBL wavelength,
there exists at least one allowed shape with the given flux.
While this procedure does not extract the full information on the EBL from the combined VHE spectra,
as instead done in Ref. \citen{costamante04,dwek05},
it provides a  general conservative upper limit.
In addition to $\Gamma_{int}\geq1.5$, an upper-limit curve for a $\Gamma\geq2/3$ blazar specrum
is also calculated,  which shifts the limits up by $\approx30-50$\%.
The latter curve corresponds to the hardest possible spectrum allowed by a SSC model 
for a monoenergetic or maxwellian electron distribution\cite{katar06,lefa11}. 

The most recent analysis based on this method is performed by Ref. \citen{meyer},
including all new detected VHE sources and taking into account the Fermi-LAT results from the first year.
The results confirm the previous findings, and are shown in Fig. \ref{meyer}.

Comparing the overall EBL upper limits with the total integrated light from galaxy counts,
it is possible to put constraints on the properties of any additional contributor to the EBL,
in particular  PopIII and low metallicity Population II stars\cite{raue09}.
Depending on the assumptions on star formation rate (SFR) and metallicity,
a limit on the SFR of the first stars of 0.3-3 \Msun Mpc$^{-3}$ yr$^{-1}$ can be derived,
in the redshift range 7-14\cite{raue09}.

\subsection{Extrapolation from Fermi-LAT spectra}
\label{extralat}
Fermi-LAT provides information on the intrinsic gamma-ray spectrum  
in an energy range (0.1-100 GeV) not affected by EBL absorption, 
for sources in the local Universe.  
The EBL can then be constrained  by assuming that the extrapolation of the Fermi-LAT spectrum 
into the VHE band is either a good estimate or an upper limit
for the intrinsic VHE spectrum of the source\cite{sanchez,orr11,georgan10}.
The rationale is that if both emissions belong to the same SED hump, the flux and spectrum at VHE
should always be lower and steeper than at HE ($\Gamma_{VHE}\geq\Gamma_{LAT}$).
In other words, the blazar spectrum should always 
be convex\footnote{In this paper  the common concept of "convex" and "concave" 
spectrum is adopted, considering the blazar emission in the SED as surface of reference.
Namely, a spectrum is convex if its photon index $\Gamma$ increases (softens) with energy 
(concavity oriented downwards in the SED), while it is concave if  $\Gamma$
decreases (hardens) towards higher energies  (concavity oriented upwards).   
Note that recently some authors\cite{meyer,hessebl,dwek2013} adopted the opposite definition 
(i.e. a convex spectrum is called  ``concave", and a blazar SED is said to be dominated 
by two ``concave" emission humps).}
The limits derived with this approach
are usually more stringent than the general spectral-hardness limits,
because the LAT spectra are on average steeper than $\Gamma=1.5$.

\begin{figure}[t]
\centering
\includegraphics[width=0.95\textwidth]{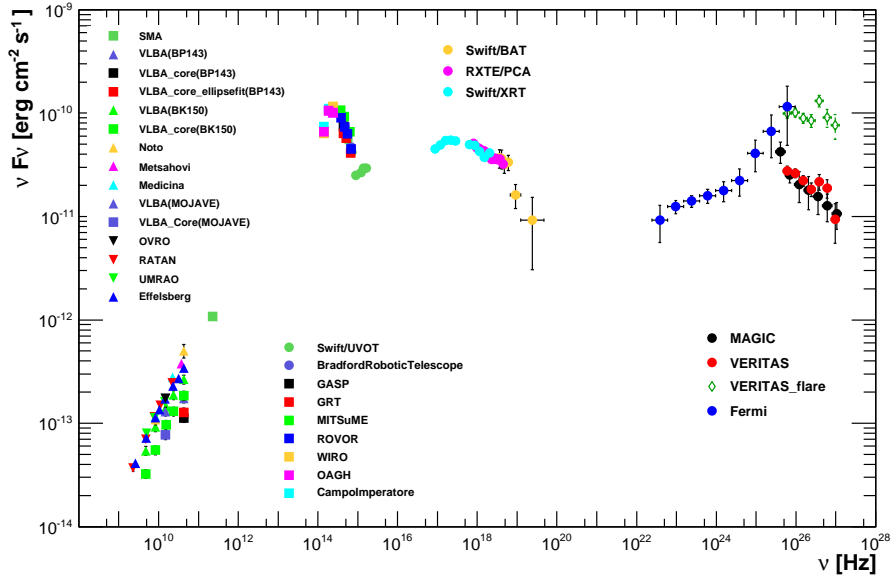}
\caption{SED of Mkn\,501 averaged over all observations taken in the multifrequency campaign 
from March 15 to August 1, 2009.  Data from different instruments, as labelled.
Optical and X-ray data have been corrected for Galactic extinction, and the VHE 
data corrected for EBL absorption using the model by Ref. \citen{franceschini08}. 
The Fermi-LAT spectrum shows that blazars can have a gamma-ray spectrum
hardening towards higher energies. From Ref. \citen{lat501flare}. }
\label{lat501}
\end{figure}

However, blazar observations show that this assumption is not well justified 
anymore\cite{gevtev}.  
On one hand,  the transition between HE and VHE bands is the energy range where blazars spectra 
tend to change the most,  since it corresponds to the peak of the gamma-ray emission.
For any given EBL model, data show that a simple extrapolation over 
more than one decade is rarely correct.
On the other hand,  blazars have shown multiple spectral components
in their synchrotron emission,  which could emerge by IC in the gamma-ray band. 
The detection of VHE spectra with $\Gamma\sim1.5-1.7$  even in case of a low EBL density
demonstrates the existence of such hard intrinsic components, 
which could possibly overcome a softer HE spectrum  yielding $\Gamma_{VHE}<\Gamma_{LAT}$. 
Over 5 decades in energy, it is not only possible but even likely that a combination 
of different emission components -- in time or in particle population or in emission mechanism -- 
can result in a concave overall spectrum.  The average Fermi-LAT spectrum of Mkn 501 (Fig. \ref{lat501})
already provides such an example\cite{lat501flare}.
EBL limits from simple Fermi-LAT extrapolations into the VHE band, therefore, 
should be considered with caution, and at most as educated guesses\cite{gevtev}.
The same applies to redshift constraints derived from assuming $\Gamma_{VHE}>\Gamma_{LAT}$
or from empirical laws of the relation between $\Delta\Gamma=\Gamma_{VHE}-\Gamma_{LAT}$ and 
redshifts\cite{prandini2010,prandini0447},  for a given EBL model. 
They are valid as far as  the source follows the assumed behavior, 
but we already know that there is a population of blazars which does not\cite{gevtev}. 

\subsection{SSC/EC modeling of the SED}
\label{sedlimits}
Another approach  --perhaps more realistic though more model-dependent--   is to try 
to predict the intrinsic VHE gamma-ray spectrum 
by modeling the full blazar's SED, with SSC or EC models.

This approach was pioneered by Ref. \citen{coppi99} in 1999, in relation to the 
additional timing information that simultaneous observations of the synchrotron and IC emission
can provide during flares.
In the SSC model,  even for a single electron population the knowledge of the synchrotron emission alone 
is not sufficient to univocally determine the IC emission. 
The number of parameters is larger than the observables\cite{tavecchio98}.
However, if the optical/X-ray and TeV variations are consistent with being produced by 
a common electron distribution, 
it is possible to reduce or even eliminate the degeneracy of the model using the timing 
information\cite{coppi99}.    If the single-zone SSC scenario works, 
it becomes then possible to robustly estimate the blazar's intrinsic 
VHE spectrum from its X-ray spectrum\cite{coppi99}. 
The analysis of almost simultaneous X-ray and VHE spectra during a high state 
of Mkn\,501 in 1997 highlighted the importance of the EBL effects in the derivation 
of the SSC parameters, and already provided relevant constraints to the EBL\cite{guy2000}.

The Fermi-LAT data can help to strongly reduce the model degeneracy as well,
by providing an anchor to the SSC/EC modeling in the IC hump.
Using all available data in the SED, and  keeping the values of the physical parameters
within the expected ``ballpark" for this kind of objects,
the result of the SSC/EC modeling is assumed as the intrinsic source spectrum in the VHE band. 
It is used either to constrain more strongly the EBL\cite{mankuz10} when the redshift is known, 
or to constrain the distance for sources of unknown redshift, 
for a given EBL model\cite{veritas1424,mankuz11}

This method is certainly more accurate than a simple power-law extrapolation. 
However, the same data provided by Fermi-LAT and the new generation of Cherenkov telescopes
show that a single-zone SSC model is not adequate and often fails to reproduce the blazars properties.

About $\sim1/3$ of all VHE-detected blazars are characterized by a hard VHE spectrum 
even with a low EBL density, locating the IC peak  at multi-TeV energies. 
The SED of these hard TeV blazars is generally not reproduced by a single-zone SSC model,
often requiring at least two particle populations to account for the full IR-to-X-ray and HE-VHE emission. 
Even in that case, the physical parameters for the hard VHE component  are very different 
(very low magnetic fields, very high particle densities and energies)
from the typical parameters of the other BL Lacs\cite{gevtev,tavecchiohbl,lefa11}.  
Furthermore, blazars show a superposition of multiple spectral components 
with different ratios of the IC to synchrotron luminosity, which can lead 
to a complex behavior during high-activity states (like gamma-ray flares in absence of
or with weak relation to synchrotron flares\cite{1959,chandranight}) 
as well as on long timescales\cite{hess2005a,hess2005b}.   

For a given HE spectrum and overall SED, therefore, the range of possible VHE spectra is
still large ($\Delta\Gamma\approx2$), even in a SSC scenario 
depending on the choice of parameters and adopted particle distributions\cite{lefa11}. 
One-zone models can work {\it a-posteriori}, but  do not work reliably {\it a-priori}.
A robust  prediction of the VHE spectrum from the HE data is not yet possible, at least for
individual sources\cite{gevtev}.

\section{Collective signal: Fermi-LAT detection of the HE absorption edge}
Fermi-LAT has detected more than 1000 blazars\cite{2LAC}
at HE over a wide range of redshifts, and often with large statistics
in the  1-100 GeV band.  In the local universe ($z<0.2$), the Fermi-LAT band is essentially
insensible to EBL attenuation, but the absorption edge shifts more and more into the LAT band
as redshift increases  ($E_{crit}(z)\approx  170(1+z)^{-2.38}$\,GeV,
where $E_{crit}$ is the critical energy  below which less than 5\% of the source photons 
are absorbed\cite{majello}, for the EBL calculation by Ref. \citen{franceschini08}.
It becomes therefore possible to try to detect the collective imprint of the first EBL 
absorption edge in the combined gamma-ray spectra of all blazars,
as a cut-off that changes amplitude and energy with redshift. 

\begin{figure}[t]
\centering
\includegraphics[width=0.6\textwidth]{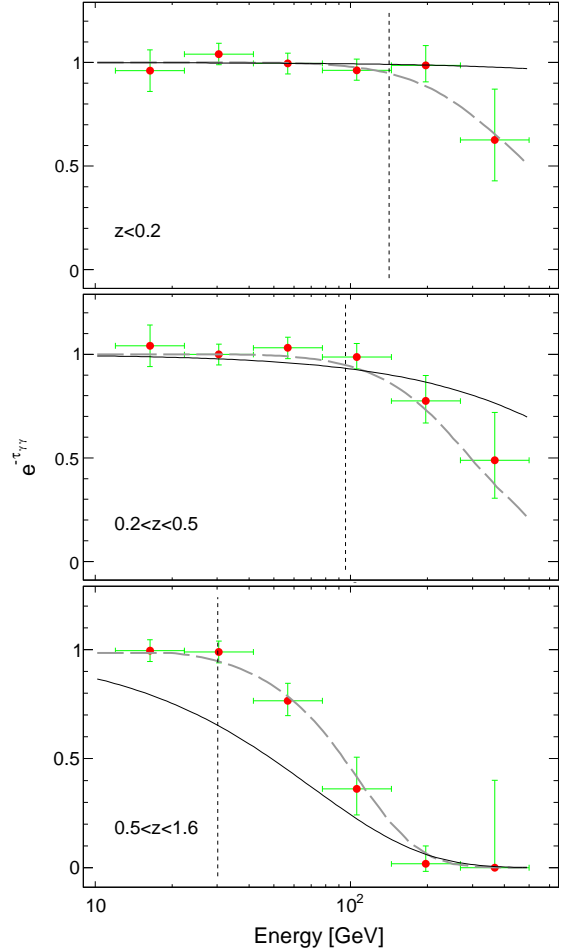}
\caption{Ratio of the average extrapolated vs observed LAT blazars spectra 
in different redshift bins, showing a cut-off feature as a function of increasing redshift.
The vertical line shows the critical energy $E_{crit}$ below which 
$<$5\% of the source photons are absorbed by the EBL, and thus where 
the source intrinsic spectra are estimated. 
The dashed curves show the attenuation expected from the EBL (with model by Ref. \citen{franceschini08}),
obtained by averaging in each redshift and energy bin the opacities
of the sample.  
The thin solid curve represents the best-fit model assuming that all the sources 
have an intrinsic exponential cut-off and that blazars follow 
the ``blazar sequence" model\cite{fossati98,gg98,ggsequence2,eileen11}.
From Ref. \citen{majello}.  }
\label{latcutoff}
\end{figure}

To this aim, Ref. \citen{majello} analyzed 46 months of LAT data in the 1-500 GeV band,
for a subset of 150 blazars with a significant detection above 3 GeV and known redshift.
Only BL Lac type objects were considered,
to minimize the possible influence of absorption at the source due to UV photons of the Broad Line Region 
(BLR, particularly strong in FSRQ\cite{anita07}).  

The sample covers a redshift range 0.03--1.6, meaning $E_{crit}$ is always $>25$ GeV.
The intrinsic spectrum of the source was then determined by fitting the unabsorbed part 
(1 GeV -- $E_{crit}$), and then extrapolated to higher energies. 
By combining all the spectra in a maximum likelihood fit, Ref. \citen{majello} 
determined the average deviation, 
above the critical energy, of the observed spectra from the extrapolated ones.

The intrinsic spectra were modeled with a log-parabolic shape in the Flux-Energy space,
and the EBL optical depth was defined as $\tau_{\gamma\gamma}(E,z) = b \cdot \tau^{model}_{\gamma\gamma}(E,z)$,
where  $\tau^{model}_{\gamma\gamma}(E,z)$ is the optical depth predicted by 
EBL models\cite{kneiske04,stecker06,franceschini08,finke10,dominguez11,gilmore12}
and $b$ is a scaling variable, left free in the likelihood maximization.
In this way it was  possible to assess the likelihood of  two important cases:
i) absence of EBL attenuation ($b$=0),  ii) the model prediction is correct ($b$=1).

\begin{figure}[t]
\centering
\includegraphics[width=0.9\textwidth]{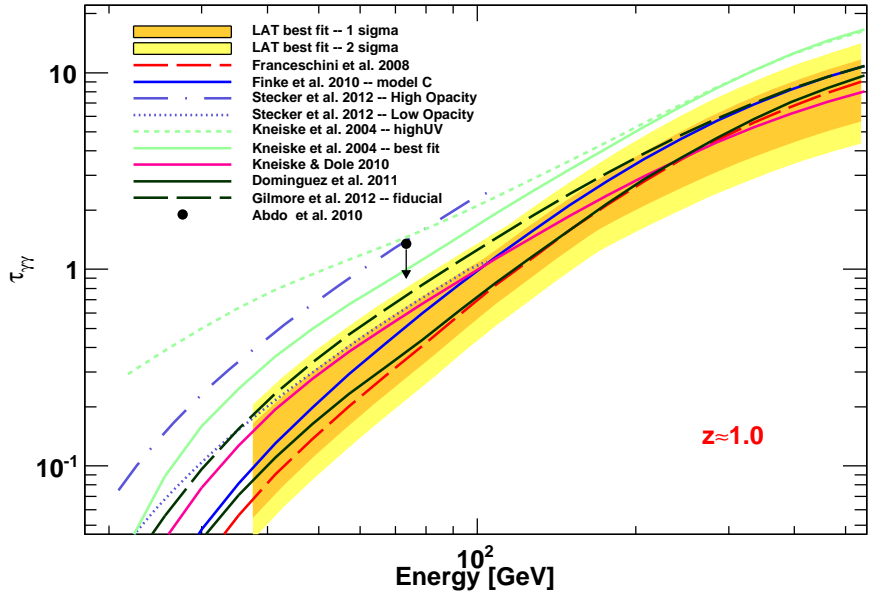}
\caption{Confidence ranges, at the 68\% and 95\% level including systematic uncertain-
ties added in quadrature, of the opacity $\tau_{\gamma-\gamma}$ from the best fits 
to the Fermi-LAT data compared to predictions of various EBL models. 
The plot shows the measurement at $z\approx$1, which is the average redshift 
of the most constraining bin (i.e. $0.5<z<1.6$). The Fermi-LAT measurement
was derived combining the limits on the best-fit EBL models.
From Ref. \citen{majello}.  }
\label{latz1}
\end{figure}

The result is the detection of an average cut-off feature in the blazar spectra,
whose amplitude and modulation in energy evolve with redshift as expected for EBL absorption
(Fig. \ref{latcutoff}).
This feature is consistent with models which prescribe a low EBL density 
(for which $b=1$ within 25\%), yielding a significance of up to $6 \sigma$.
Models with a larger EBL intensity, particularly in the UV\cite{kneiske04,stecker06,stecker12},
would produce a stronger-than-observed attenuation feature and are therefore 
incompatible with the Fermi-LAT data\cite{majello}  (Fig. \ref{latz1}).

The main problem is to assess if this spectral cut-off is not intrinsic to the gamma-ray sources,
in total or in part.  In fact, the BL Lac sample is not homogeneous, but changes 
composition with redshift, being dominated by High-energy-peaked BL Lacs (HBL) at low redshift 
and Low-energy-peaked BL Lacs (LBL) at higher redshifts ($z>$0.5).
LBL are characterized by steeper gamma-ray spectra and by an electron distribution
with lower maximum energy, as well as a possible (though weak on average) BLR\cite{giommi12}.  
In principle, therefore, the cut-off could be caused by the end of the particle distribution 
or some level of internal absorption. 
However, if intrinsic, the cut-off is expected to evolve differently with redshift than is 
observed (see Fig. \ref{latcutoff}), assuming that blazars as a population are 
indeed distributed according  to the so-called 
"blazar sequence\cite{fossati98,gg98,ggsequence2,eileen11}". 

Regardless, the agreement between the intensity  of the UV background as measured 
with Fermi-LAT\cite{majello}  
and that due to galaxies individually resolved by the Hubble Space 
Telescope\cite{gardner2000,madaupozzetti}
leaves little room to an intrinsic origin of the cut-off feature, 
as well as to any residual diffuse UV emission. 
If intrinsic, the EBL attenuation must be even lower than the already low predictions 
from galaxy evolution models ($b<1$).
A lower-than-expected EBL (lower than the galaxy counts limits) would require
modifications  of fundamental physics laws (see Sect. \ref{exotic}).
A higher-than-expected EBL would be compatible with the Fermi data only if blazars 
on average have a second harder component in their intrinsic spectra emerging at higher energies, 
which shifts with redshift as to perfectly mimic the feature evolution of a lower-density EBL.
Such extreme fine-tuning is difficult to justify in absence of any other evidence.

The average cut-off feature discovered with Fermi-LAT, therefore, is most likely the first direct detection 
of the gamma-ray horizon, and a further evidence of the gamma-gamma absorption process 
caused by a low-density EBL.
The Fermi measurement constrains\cite{majello,raue09,gilmore11} the redshift of the maximum
formation of Population III stars to be at $z\geq10$, 
and its peak co-moving star-formation rate  
to $\leq0.5$ M$_{\odot}$ Mpc$^{-3}$ yr$^{-1}$.

\subsection{Collective signal: application to the VHE data.}
With a sufficient number of blazars and photon statistics,
it becomes feasable to try to measure the collective imprint of the EBL
also on all VHE spectra of blazars together.
This study was recently performed on the H.E.S.S. data\cite{hessebl},
by fitting simultaneously the EBL optical depths 
and the intrinsic spectra together, with a maximum likelihood method.
The underlying assumption  is that the true intrinsic$+$absorption spectrum combined
should maximize the likelihood.

Only objects detected at least above 10$\sigma$ were considered.
To reduce the space of free parameters to a manageable size, 
the EBL spectrum was fixed to a single shape 
(adopting the calculation by Ref. \citen{franceschini08} as template), 
leaving only the normalization free.
The blazar intrinsic spectrum was assumed to be smooth (i.e. with no abrupt breaks or pile-up features),
and overall convex, but   
with no assumption on its hardness.
The spectra  were fitted with a single powerlaw or log-parabola model, 
with or without the addition of exponential and super-exponential cut-offs. 
Among all fits, the intrinsic model giving the highest \rchisq\ probability was chosen 
for each source. The overall test statistic (TS) as a function of the  EBL optical depth 
normalization was then computed, combining the TS functions of all the separate sources and datasets. 
The result  is a ``highest-TS  measurement" of the EBL normalization at 1.4 \m\ of 
$\lambda F_{\lambda}=15\pm2_{stat}\pm3_{syst}$ \nw. This value is  $\sim8-9\sigma$ above 
the null hypothesis  (i.e. absence of EBL absorption) and $1.8\sigma$ above the unscaled 
template of EBL model by Ref. \citen{franceschini08},
but within the upper limits  derived from the hardness of the spectrum\cite{hessebl}.

There are two main problems with this approach, 
besides the weakness of the assumption of only convex blazar spectra 
(see Sect. \ref{extralat} and \ref{sedlimits}).
One is on the application of the maximum likelihood method itself,
the other is the bias introduced by combining the TS functions.

Given the complex interaction between the intrinsic blazar spectrum and the shape of EBL attenuation, 
there is no guarantee  that the combination of the two 
which maximizes the likelihood is the correct one. 
Indeed, among different objects  the maximum probabilities occur at very different EBL normalizations, 
implying {\it de facto} that the true value is certainly {\it not} given by the peak of the \chisq\ probability 
(see Fig.1 in Ref. \citen{hessebl}). This invalidates the main assumption. 
In addition, in each single source  
the \chisq\ probability peaks at very different opacity levels depending on the intrinsic models,
but often  with very similar peak probabilities (e.g. between 50 and 60\%)
for very different opacity values,   
making the choice of one of them somewhat arbitrary. 
The reported uncertainties, therefore, appear strongly underestimated, undermining
the significance of the detection.
Fact is, all VHE spectra detected so far  can be well fitted with smooth functions and
perfectly acceptable statistical goodness (\rchisq $\lappr 1$) for any EBL normalization  
between the lower limits from galaxy counts (and even below that) 
and  the direct measurements in the 2-4 \m\ band.

Concerning the best-fit value, the combination of the TS functions makes the result, by construction,
heavily biased towards datasets with  higher statistics,
which in this case means  the datasets of PKS\,2155-304\cite{hessebl}.
This object is characterized by a rather steep and intrinsically curved spectrum, 
which partly cancel  the amplitude of the flattening feature imprinted by the EBL
on the observed spectra. 
This allows for a higher-than-real  EBL density to be considered viable with a high TS.
As a result, the combined TS is likely biased towards higher EBL values than they actually are.

Despite these shortcomings, this result does confirm that 
on average the blazar VHE spectra seem better fitted including some amount of EBL absorption 
rather than not, thus supporting  the Fermi-LAT result against the null hypothesis.
This is precisely the expected effect of the  flattening feature shown in Fig. \ref{flattening}.  
It is always possible to consider this result as due to 
an intrinsic property of the blazar emission, but again it would require an extreme fine tuning to make it 
always appear at the same energy, irrespective of source redshift, as to mimic precisely the effects 
of absorption by an EBL spectrum consistent with  galaxy evolution.

\section{Alternative scenarios}
Alternative explanations 
look for ways to circumvent the severe intergalactic absorption 
caused by photon-photon collisions  with the EBL photons,
in order to increase the {\it effective} mean free path of TeV gamma-rays.
This can be accomplished  either by introducing new fundamental physics,
or by considering the observed gamma-rays as secondary emission  
resulting from the development of electromagnetic cascades in the intergalactic medium.

\subsection{New exotic physics}
\label{exotic}
A ``straightforward" way to increase the transparency of the universe 
is to simply modify the cross section of the photon-photon collision and pair production process.
Quantum-gravity (QG) theories predict in general a breakdown of familiar physics when
approaching the Planck energy scale, and a violation of
Lorentz invariance arises naturally in various such theories\cite{camelia,jacob}.
The modification of the photon dispersion relation, such that vacuum becomes
a refractive medium with index $\simeq 1+(E/M_{QG})^n$ ($n$=1 or 2), 
can lead to a shift of the threshold  for pair production at higher energies,  
as well as an energy-dependent  time-of-flight for photons.
The shift of the gamma-gamma cross section  is expected to cause the optical depth
to decrease again as photon energy increases, leading to a re-emergence of the source flux 
at higher energies which can be best tested with the future Cherenkov telescope
like CTA\cite{cta} and HAWC\cite{hawc}.

Time-of-flight analyses provide the most generic method of testing
Lorentz Invariance\cite{camelia}. They have been performed for different cosmological gamma-ray sources,
looking for delays in the time of arrival of gamma-ray photons at different energies
during the fastest flares. 
Though the detection of such feature needs to address the possibility
of its astrophysical origin (i.e. intrinsic to the source), upper limits on the delays
allow lower limits on the QG mass scale to be derived (see review\cite{bolmont11}).
The most stringent limits have been derived so far on the linear term,
from gamma-ray bursts with individual photons above 10 GeV  
($M_QG>1.5\times10^{18}$ GeV  from GRB\,080916C\cite{lat_grb1} 
and $>1.5\times10^{19}$ GeV from GRB\,090510\cite{lat_grb2})
and from the brightest TeV blazars at VHE energies ($M_QG>2\times10^{18}$ GeV from 
PKS\,2155-304\cite{hess_qg1,hess_qg2}, and $M_QG>3\times10^{17}$ GeV from Mkn\,501\cite{magic_qg}).

Another way to increase the effective mean free path is to assume the oscillation of photons
into  hypothetical axion-like particles (ALPs) in the presence of 
a magnetic field\cite{deangelis07,simet,alps_review2010}.
No evidence  has been found in support of this possibility so far.
One of the most constraining bounds on the the product of the photon-ALP coupling 
$g_{a\gamma}$ times B has been obtained  by studying the rotation of the linear polarization 
measured in UV photons in radio galaxies at cosmological distances 
(see Ref. \citen{horns_alps} and references therein).
It is found that $g_{a\gamma} B  \lesssim 10^{-11}$ GeV$^{-1}$ nG for ultralight 
ALPs with mass  $<10^{-15}$ eV.

\subsection{Cascades from ultra-high-energy  protons}
Cosmic ray (CR) protons with ultra-high energy (UHE) below the GZK cutoff 
(e.g. $E\sim 10^{17}-10^{19}$ eV) 
do not lose a significant part of their energy in interactions with background photons.  
They can thus provide an effective way to transport energy over large cosmological distances,
to be deposited back into photons closer to the observer via electromagnetic cascades.
 
Cascades are initiated by interactions with the cosmic microwave background (CMB) and EBL photons via 
pair production $p\gamma  \rightarrow p e^+ e^-$ 
and photomeson reactions  $p+\gamma_b  \rightarrow  p + \pi^0$, 
and are supported by the IC scattering of electrons on the CMB and $\gamma-\gamma \rightarrow e^+ e^-$
interactions again with EBL and CMB photons. 

As long as the the magnetic field along the path is small enough
($B\lesssim 10^{15}$ Gauss),  UHE protons can travel almost rectilinearly and 
the broadening of both the proton beam and the cascade electrons due to deflections in the IGMF 
can be less than the  point spread function of the Air-Cherenkov detectors ($\sim 0.1\degree$). 
Secondary gamma-rays can then add up to the primary emission without smearing
the VHE point-like images of blazars. 

Secondary gamma-rays are produced all along the path, but 
because they are generated closer to the observer, 
they are correspondingly less suppressed by EBL absorption.
Though the probability of the $p\gamma$ interaction is very small, for distant sources ($z>0.1$)
and sufficiently high proton flux,   the secondary emission can become dominant
over the primary emission, the latter being filtered out by the stronger EBL suppression
over the full distance\cite{essey1}.

This scenario presents  distinct spectral and temporal features. 

{\bf a)} The secondary spectrum  does not depend on the details 
of the proton spectrum or maximum proton energy (as long as E$_{max}>10^{18}$ eV\cite{essey2,felixz1}),
but only on the source redshift.  For a given distance and EBL spectrum, therefore, the power
emitted in cosmic rays is the only free parameter which can be used to fit the data.
Therefore all sources with similar redshift should have very similar spectra.
At $z=0.14$ (e.g. for 1ES\,0229+200\cite{essey2}), the spectral index from secondary emission 
is calculated around $\Gamma\sim2.5$  in the 1-10 TeV range and harder below 1 TeV
(e.g. for 1ES\,0229+200\cite{essey2}),
reaching around $\Gamma\sim4.1$ at $z\approx0.4$ in the 0.1-1 TeV band 
(e.g. for 3C\,66A\cite{essey1,essey2}). 

{\bf b)} No temporal variability is expected from secondary gamma-rays on short timescales.
Proton deflections and the electromagnetic cascade introduce long time delays,
of the order of a year for 1 TeV photons 
(and longer at lower energies), already at z=0.17\cite{prosekin}.
Any source variation on short timescales (days-weeks) should therefore be washed out
\footnote{The only exception is if the IGMF is indeed very low ($B\lesssim 10^{-17}$ Gauss)
and for gamma-rays above 10 TeV (see e.g. Ref. \citen{prosekin}).}.
Furthermore, gamma-ray variations should not be correlated with  
variations at lower frequencies (e.g. in optical or X-ray bands), which are 
related to the primary emission. 

This scenario faces two main problems.
 
{\bf 1) Energetics}. Even if the UHE protons are beamed at the origin
like the primary gamma-rays from blazars, 
for all possible sites of CR production in AGN    
(black-hole, jet, lobe, terminal shocks etc.)
they have to travel for 
kpc through  regions  of relatively strong magnetic field,
from milli to nano Gauss (e.g. the jet itself, the lobe, possibly the host galaxy and its surroundings, 
the possible cluster where the AGN lives and the large-scale filament;
see e.g. Ref. \citen{murase12}). 
A magnetic field of 1 nG over 1 kpc is sufficient to deflect a $10^{17}$eV  proton 
for more than $10\degree$,
and  1 $\mu$G is sufficient over less than 10 pc. 
The net effect is either to  strongly broaden
the proton beam (if B is random on sufficiently small scales), or to deflect it
away from the observer (if B is coherent over a sufficiently large region).
Therefore, the jet opening angle cannot be assumed for the proton beam as well, 
in order to reduce the requirements on the intrinsic cosmic ray luminosity 
(e.g. by a factor $10^{-2}$ as in Ref. \citen{essey1}). 
For realistic estimates, blazars and AGN should be treated at minimum as isotropic sources of UHE protons.
The luminosities in CR required in this scenario are then
of the order of $10^{46}-10^{50}$ erg/s \cite{essey2,felixz1},
for sources (TeV BL Lacs) whose total jet kinetic power  --considering one proton for electron--
is estimated in the range $10^{42}-10^{46}$ erg/s \cite{celottigg_powers}.  

{\bf 2) Deflections in the large scale structure}. 
This model requires that a typical line of sight does not cross regions of magnetic field
larger than $10^{-15}$ G.  While the size and space density of scatterers like galaxies and galaxy clusters
in the Universe is small enough\cite{berezinsky,felixz1}, the impact of the filaments in the 
large scale structure (LSS) is still uncertain. 
Observations and simulations shows that the Universe is a large
mesh of voids of size 20-100 Mpc, connected by filaments\cite{bolshoi,bolshoi_dm,sdssIII,einasto12}. 
The intensity of the magnetic field seems to follow the mass density as $B\propto \rho^{2/3}$, 
yielding nano-Gauss fields over possibly 10-20 Mpc regions\cite{bruggen05}. 
Though the size, volume filling factors 
and geometry of the magnetic regions remain highly uncertain, it seems unlikely for a random line of sight 
to Gpc distances not to cross any region at near-nanoGauss level. 
There is no proper study so far on  the probability that a random line of sight
intercepts regions with $B>10^{-15}$ Gauss, and how this probability increases with redshift. 
However, it is safe to assume that, by its very nature, this scenario should work only for a small subset of 
line of sights (i.e sources), and not for all blazars.

\subsubsection{Comparison with observations}
The exact redshift and energy of the transition between the primary and secondary spectrum
is uncertain, being also source-dependent. 
It  is therefore important to assess the viability of this scenario 
in the whole energy range above $E>0.1$ TeV, for sources at $z>0.1$.


Considering spectra up to 1-2 TeV, 
which form the large majority of all VHE detections so far, 
blazars show a wide range of spectral slopes in the same redshift bins, up to z$\sim$0.3:
for example, RXJ\,0648+15\cite{veritas0648} ($\Gamma=4.4\pm0.8$, z=0.179) 
and 1ES\,1101-232\cite{nature_ebl,hess1101} ($\Gamma=2.9\pm0.2$, z=0.186).
The VHE emission is often strongly correlated with variations at lower energies (X-rays, optical). 
In principle only the hard VHE sources should be those dominated by the secondary emission,
while the soft and variable spectra would be explained by the primary emission. 
However,  day-scale variability is seen also in hard sources (e.g. 1ES\,1218+304\cite{veritas1218b})
and the line of sight of another hard blazar  (H\,2356-309\cite{hess2356,hess2356mwl,correct2356},
$\Gamma=3.1\pm0.2$, z=0.165) 
is known  to intercept at least 3 LSS filaments\cite{fang,zappacosta}  at z=0.03 (Sculptur Wall), 
z=0.062 (Pisces-Cetus supercluster) and z=0.128 (Farther Sculptur Wall).
These results disfavour a secondary origin of the gamma-rays  and 
demonstrate that hard features are a property of the primary emission as well.

Furthermore, the predicted secondary spectrum fits well the  hard TeV sources
only in the case of the Stecker 2006 EBL model\cite{essey2}.
The latter is however excluded by the Fermi-LAT result and is at odds with all
most recent EBL calculations.
Changing EBL values, the predicted spectra do not match well the VHE data anymore 
(see Fig. 1 in Ref. \citen{essey2}).
For blazars at z=0.4-0.5,  the measured spectra would appear consistent  
with secondary emission (e.g. 3C\,66A\cite{essey2}, z=0.4 $\Gamma=4.0\pm0.5$).
However, some sources with compatible spectra do show fast variability correlated with 
X-ray and optical activity  (e.g. 1ES 1553+113\cite{atel4069,danforth1553}, $\Gamma\approx4.4\pm0.2$).

In summary, up to $z\sim0.5$ and E$\sim$1-2 TeV, 
observations do not support --in fact mostly disfavour-- the interpretation of gamma-rays 
as secondary emission from UHE protons.  
This is also in agreement with some estimates\cite{essey2,prosekin,murase12} 
for the transition between primary and secondary emission (E$_t\geq$1 TeV).
Since the EBL limits around few microns are based on VHE spectra up to these values 
of energy and redshift,
the scenario of proton-induced cascades seems not able to relax the constraints on the EBL 
at the peak of the stellar contribution.

Above E$>$1-2 TeV, instead, the issue remains open. 
At present, only three distant blazars 
have been measured up to 6-10 TeV:  PKS\,2155-304\cite{chandranight,2155overall} (z=0.116), 
1ES\,1426+428\cite{hegra1426} (z=0.129)
and 1ES\,0229+200\cite{hess0229} (z=0.140).  
The fast, correlated variability and spectral steepness of PKS 2155-304 above 1 TeV\cite{chandranight}
rules out a secondary origin for the gamma-rays 
in this source.
This leaves 1ES\,0229+200 ($\Gamma\sim2.5\pm0.2$) and 1ES\,1426+429 ($\Gamma\sim2.6\pm0.6$) 
as the {\it only} two sources for which the cascade scenario  is actually  viable,
though their spectra are fully consistent with --and naturally explained by-- 
the characteristic feature imprinted by pure EBL absorption
onto a primary emission (Fig. \ref{flattening}).
 
Though this scenario is not favoured by the data so far,  
it is important to keep in mind  that  the efficiency of energy transfer 
from protons to gamma-rays reaches its maximum at z$\sim0.1-0.3$ (depending on gamma-ray energy),
and that it can be rather high ($>1\%$) even at 10 TeV\cite{felixz1}.
This scenario therefore provides a channel for the production of multi-TeV gamma-rays 
that in principle could  give a non-negligible contribution to the quiescent component 
of the VHE emission from some nearby blazars\cite{felixz1}.

\subsubsection{TeV gamma-rays beyond z=1}
The situation is completely different for sources at large redshifts. 
Beyond z=1, gamma-rays above several hundred GeV
are strongly suppressed by intergalactic absorption, with optical depths
$\tau_{\gamma\gamma}\gtrsim 10$ for any realistic model of 
the EBL\cite{franceschini08,dominguez11,gilmore12}.
Only a negligible fraction of the primary emission can therefore survive 
and reach the observer (unless considering apparent gamma-ray 
luminosities well in excess of $10^{50}$ erg/s).
 
The detection of TeV gamma-rays from objects beyond z=1, therefore, would 
challenge standard interpretations, and would justify 
the introduction of changes in fundamental laws of physics   
like the violation of Lorentz invariance
or oscillations with ALPs particles.
To this respect, the scenario of secondary photons from UHE protons
accelerated at the source would become very appealing.
This model could provide the {\it only viable astrophysical interpretation}
consistent with conventional physics\cite{felixz1},
provided that the spectral and temporal behavior of the data are consistent, 
and that energetics and line-of-sight problems can be overcome.
Remarkably, it would directly imply that AGN jets are very powerful and very efficient 
accelerators of UHE protons.

Such scenario was indeed fostered by an initial claim of 
large redshift ($z\geq1.246$) for the TeV blazar PKS\,0447-489\cite{hermi0447}, 
whose gamma-ray spectrum was measured up to $\sim$1 TeV\cite{hess0447}.
The VHE spectrum was found consistent with secondary emission\cite{felixz1},
and provided a case study for such possibility.
However, recently it has been shown that  the redshift estimate is most 
likely incorrect\cite{fumagalli,pita}.
The absorption feature seen at  6280 \AA\ is not a redshifted Mg II line (as initially thought)
but most likely originates in Earth's atmosphere, being detected in the spectra of 
multiple standard stars in the same observing runs\cite{fumagalli}.
The redshift of PKS\,0447-489 remains thus uncertain.

Given the importance and far-reaching implications of a detection of TeV gamma-rays 
beyond z=1, dedicated observational campaigns  on TeV candidates with established large redshift 
seem fully justified,  already with present Cherenkov arrays  
but especially with upcoming facilities like CTA\cite{cta}, characterized by
higher sensitivity and larger effective areas in the multi-TeV range.

\begin{figure}[t]
\includegraphics[width=0.48\textwidth]{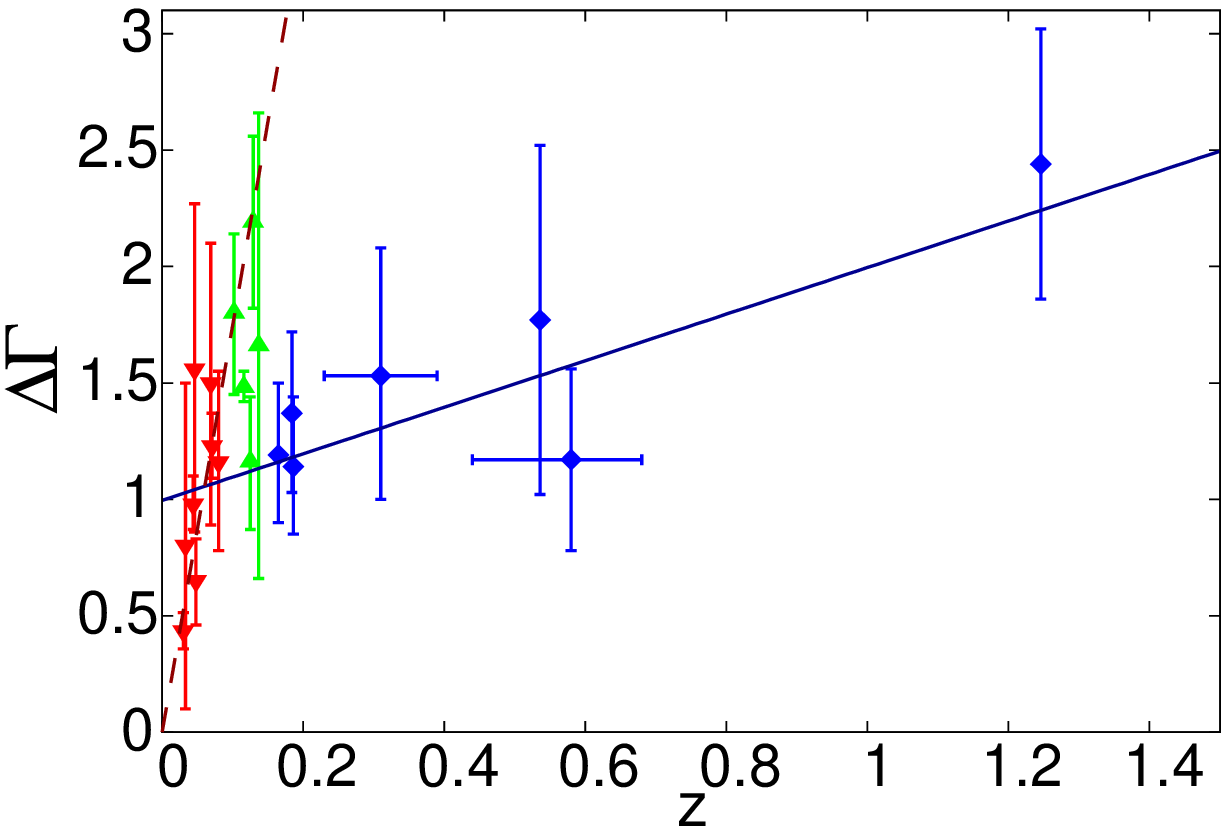}
\includegraphics[width=0.51\textwidth]{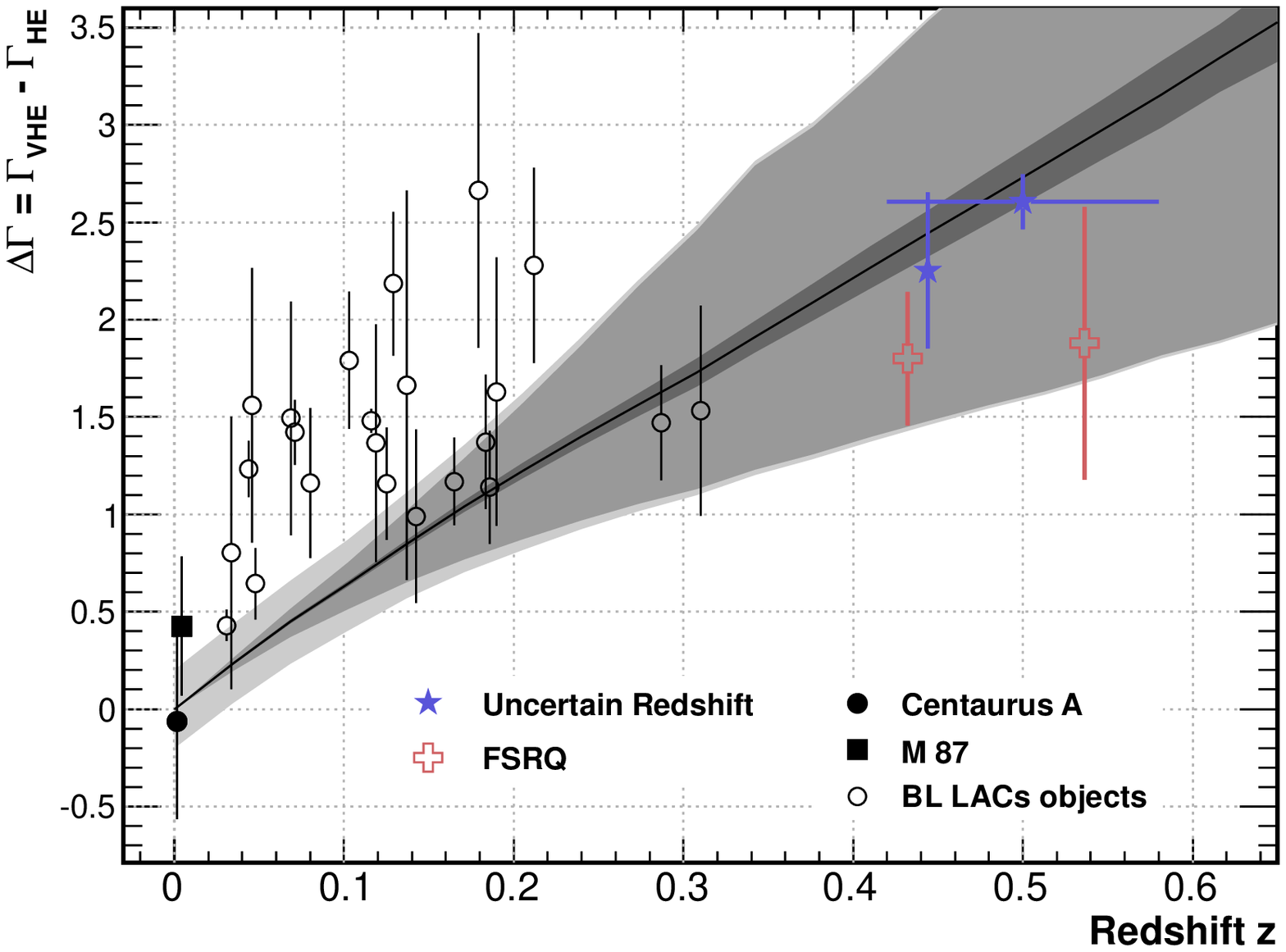}
\caption{Spectral change $\Delta\Gamma\equiv\Gamma_{VHE}-\Gamma_{HE}$
as a function of redshift for TeV-blazars detected by Fermi 
(data from the second Fermi catalog\cite{2LAC}).
Left: tentative interpretation of the existence of two populations,
based on the wrong redshifts of the two most distant objects 
and the (wrong) expected relation due to EBL absorption\cite{stecker06,steckerscully10}.
From Ref. \citen{essey2}.
Right: more correct evolution of the $\Delta\Gamma$ break theoretically 
expected for a power-law spectrum 
by pure EBL absorption with the  EBL model by Ref. \citen{franceschini08}.
Grey bands show the uncertainties  estimated from the limited energy resolution of Cherenkov telescopes, 
different threshold energies and systematic errors (0.2). From Ref. \citen{sanchez}.
FSRQs are also shown for illustration, but should not be considered in this plot because
their HE and VHE emission most likely come from different zones and epochs. 
In this respect, note that also in BL Lacs 
$\Delta\Gamma$ could be intrinsically negative by 0.5 or more.}
\label{deltagamma}
\end{figure}

\subsection{On the $\Gamma-z$ and $\Delta\Gamma-z$ relations}
Since EBL absorption increases with distance,  
both the observed VHE photon index $\Gamma_{VHE}$  of the detected sources 
and the difference in photon index between
the unabsorbed (HE) and absorbed (VHE) 
parts of the gamma-ray spectrum  ($\Delta\Gamma\equiv\Gamma_{VHE}-\Gamma_{HE}$)
are expected to increase with redshift.
This is indeed observed, but it is often claimed that the dependence with redshift
appears weaker than expected by pure EBL absorption\cite{agncta,deangelis09,essey_weakz},
indicating the need for alternative scenarios.

However, it is important to realize that 1) blazars show a wide range of intrinsic spectra,
which vary both in time and among different objects.
Intrinsically, the $\Gamma_{VHE}$ can range between 1.5 and 3-4, according to their SED.
Furthermore, 2) for a fixed intrinsic spectrum the observed (absorbed) spectral index
takes different values (even by $\Delta\Gamma_{VHE}\sim1$) depending on the detected energy band, 
due to the energy dependence of the EBL optical depth.
Both these effects cause a large scatter in the $\Gamma-z$ plot, which is also possibly biased
by source luminosity and target choice. A fit in the $\Gamma-z$ space, therefore, 
is not a valid method to derive EBL conclusions.  
The important observable  is the envelope of the  measured spectral slopes,
which is delimited and drawn by the hardest spectra observed at each redshift.
This envelope shows indeed a clear evolution with redshift (see Fig. 4 in Ref. \citen{agncta}) 
which is fully consistent with the expectations for a low EBL model\cite{costamante_barca}.

Similar considerations apply to the $\Delta\Gamma-z$ dependence,
which was used to claim the existence of two populations of VHE spectra\cite{essey2}.
The evidence for two populations was mainly based on the initial estimates of high redshift 
for the two TeV blazars PKS\,0447-489 (z=1.246) and 3C\,66A (z=0.58), 
and in comparison with the very strong $\Delta\Gamma-z$ evolution 
predicted\cite{steckerscully06,steckerscully10}  
for the high-EBL model by Ref. \citen{stecker06} (Fig. \ref{deltagamma}).
The latter however is ruled out by both Fermi-LAT and VHE results.
The evidence disappears completely considering the updated redshift estimates 
(uncertain for  PKS\,0447-489\cite{fumagalli,pita}  and $z<0.41$ for 3C\,66A\cite{furniss13})
and the milder redshift evolution given by more realistic EBL models consistent 
with Fermi-LAT results\cite{sanchez} (see Fig. \ref{deltagamma}).
Furthermore, to appear in the plot  a source needs to be well detected in both HE and VHE bands.
As redshift increases, more luminous sources are detected, which tend to be characterized 
by lower-energy peaked SEDs and thus softer Fermi-LAT and intrinsic VHE spectra.  
Since it is difficult for Cherenkov telescopes to detect faint VHE spectra with slopes much
steeper than $\Gamma\sim4.5-5$, $\Delta\Gamma$ cannot show values much larger than $\sim2.5$.
This introduces an observational limit in the  $\Delta\Gamma-z$ plot, as in fact observed.
Because of these biases, and the possibility of the VHE spectra to be harder than the HE ones,
it can be expected for  $\Delta\Gamma$ to exhibit  
a weaker dependence with redshift than $\Gamma_{VHE}$.
Again, so far the observational data are fully consistent with a low EBL scenario.

\subsection{A possible ``pair-production anomaly" ?}
Considering the lowest possible EBL model as given by the galaxy counts
in each waveband\cite{kneiske10}, a tendency has been recently noted\cite{horns_meyer} for the 
absorption-corrected VHE spectra to show an upturn 
in the transition region from the optically thin to the optically thick regime
(from $\tau<1$ to $\tau>2$). For the 7 sources with data in both regimes, 
this transition occurs on a broad range of energies (0.4-21 TeV) and redshifts (z=0.031-0.539),
yielding a combined significance between 2 and 4$\sigma$ depending on the consideration 
of different systematic effects\cite{horns_meyer}.
Since intrinsic source features would imply an extreme fine tuning,
this trend has been interpreted as indication of a suppression of the pair production
in the propagation of VHE photons, possibly due to ALPs coupling.

However, the optically thick data points are typically 
at the endpoint of the measured energy spectra, where the statistics is very low
and the limited energy resolution leads to possible spill-over effects\cite{horns_meyer},
which can cause a systematic over-estimate of the flux. 
Furthermore, the lower-limit EBL model might not represent a ``minimal case"  
for these studies.
The correction for absorption depends strongly also on the shape of the EBL spectrum.
If the lower-limit EBL model underestimate the true EBL more strongly at $\sim1\mu$ than 
at $\sim10\mu$, for example, it would lead to a flatter-than-real EBL spectrum between the two wavelengths 
and thus to an apparent excess or upturn in the reconstructed spectra in the 1-10 TeV range, where 
most of the data endpoints actually are. Above 10 TeV, the reconstructed spectra are very sensitive 
to even tiny changes in the EBL shape in its warm-dust component, which is highly uncertain.
In fact, the anomaly seems to disappear as soon as the statistics increases
and with a more standard EBL model as given by Ref. \citen{franceschini08} 
(e.g. for the 2006 PKS\,2155-304 flare spectrum\cite{chandranight}, 
detected up to 6-7 TeV corresponding to $\tau\sim2.5-3$).
Together, the low statistics (both in photons and number of sources),
the data-analysis uncertainties,  the EBL-shape effects 
and the lack of anomaly with higher-statistics spectra and more standard EBL models
point towards a spurious origin for the anomaly
rather than a change to the gamma-gamma cross-section. 

\section{Blazars, EBL and the intergalactic magnetic fields}
The energy of the primary gamma-ray is not lost in the interaction with EBL
photons. The resulting electron-positron pair can upscatter CMB photons
again into gamma-ray energies, as they cool,  
with typical values $E_{\gamma}\sim 1 (E_0/40 {\rm TeV})^2$ TeV\cite{halos1}.

For photons above $\sim20-50$ TeV,
the photon absorption length $\lambda_{\gamma\gamma}(E)$ is typically short,
the pairs are produced within few Mpc of the source, in regions where the magnetic field is 
still relatively high, and thus cascading effectively surrounds the source with a giant, 
isotropically emitting halo of pairs\cite{halos1}.
If the gamma-ray source is a strong emitter above 20-50 TeV, therefore, 
the pair halo radiation can become observable at VHE below few TeV, 
as extended emission whose size depends essentially on the local EBL intensity\cite{halos1,anant09}.
No imaging of the pair halos has been obtained so far, though hopefully
their emission could be within reach of the future Cherenkov telescope CTA\cite{cta}.

For photons below $\sim10-20$ TeV, instead, the pairs are produced predominantly in the 
intergalactc space
and the secondary emission occurs in the GeV domain\cite{neronov09}.
If the IGMF is large enough, pairs are deflected significantly before
cooling by IC, broadening and suppressing the secondary emission flux.
If it is small enough, the secondary emission can remain concentrated close to the primary emission cone,
thus adding to the object's HE spectrum.
The comparison of the HE with the VHE spectra (which determines with the EBL the absorbed energy flux),
can therefore probe the poorly known  strength of the IGMF\cite{neronov09,neronov10}.
The Fermi-LAT spectra and upper limits obtained so far on several  TeV BL Lacs
allow a lower limit to the IGMF intensity to be put 
at a level of $B\gtrsim 10^{-17}$ G\cite{neronov10,tavecchio10,dolag11,dermer11,taylor11}.

The limits on the IGMF depend however on the intensity of the EBL. 
A higher EBL leads to a higher absorbed gamma-ray luminosity, thus generating a stronger cascade 
contribution  which require a stronger IGMF not to overproduce the Fermi-LAT measurements
at HE\cite{vovk0229}.
On the other hand, a higher EBL implies a harder intrinsic source spectrum,
leaving more room in the HE band to accomodate the additional softer cascade emission. 
The recent detection of 1ES\,0229+200 with Fermi-LAT\cite{vovk0229} allowed the exploration of the 
3D parameter space given by IGMF strength vs EBL intensity 
(using template by Ref. \citen{franceschini08}) 
vs intrinsic photon index $\Gamma_{int}$ (varied within the range from 1.5--0\cite{vovk0229}).
Taking into account the Fermi-LAT result\cite{majello} in favour of a low EBL density,
which means $\Gamma_{int}\gtrsim1.5$, the 1ES\,0229+200 data at HE and VHE together
constrain the IGMF to $B\gtrsim 10^{-16}$ G.
This constrain could be even stronger by assuming that the source was stable at VHE on time scales 
longer than the $\sim$3 years  timespan of the H.E.S.S. and VERITAS 
observations\cite{dermer11,hess0229,murase12}. 
This IGMF lower limit is essentially determined by the statistical error on the Fermi-LAT data,
since the HE spectrum should be strongly dominated by the primary emission, as also indicated
by the variability of the HE detection 
(the bulk of the Fermi-LAT signal comes from 2011 observations\cite{2LAC}).

\section{Conclusions}
The new data provided by Cherenkov telescopes and the Fermi satellite 
have brought substantial progress in the study of the EBL through gamma-rays.
Together with the lower limits from the integrated light from galaxies, 
the VHE spectra of hard BL Lacs and 
the collective signal in the Fermi-LAT band  have dramatically narrowed down the possible
region in the parameters space for a consistent picture of both EBL spectrum, galaxy evolution 
and overall blazar phenomenology.
The consistent and convergent picture which is emerging from all the different data 
calls for a low-density EBL, as produced by standard galaxy evolution models.  
From UV to MIR wavelengths, the EBL is now pinned down in its direct starlight component
to a level better than 10-15\%.  Consequently, 
also the average intensity of the IGMF is constrained to $B\gtrsim 10^{-16}$ G.

At the time of writing, {\it all} gamma-ray data from blazars 
are fully consistent with  this simple picure, 
and no significant evidence has clearly emerged  
either requiring or suggesting  more complicated or exotic explanations.

Given the small room left between upper and lower limits,  
more and better data are expected to provide  confirmations but no further improvements 
in the EBL stellar hump. The limiting ``wall" is essentially represented by the systematic uncertainty 
given by our ignorance in the details of blazar physics and jet physical conditions\cite{preston,felixz1}. 
In this respect, the precision with which some blazar models could in principle
predict VHE spectra might be deceptive, being far from a comparable accuracy.

The most important progress and results are likely to come 
from gamma-ray spectra in the $\sim1-100$ TeV range, as allowed for example
by the upcoming small-telescopes array in CTA\cite{cta}, its precursor ASTRI project\cite{astri},
and the HAWC observatory\cite{hawc}.  
Multi-TeV spectra can open new scenarios,  providing key answers to the study of
1) fundamental physics aspects, possibly revealing TeV gamma-rays from high redshift sources,
2) blazars as possible sources of UHE protons; 3) extreme acceleration and emission processes
in the hard TeV blazars, locating the peak of the IC emission in the SED above 1-10 TeV, and
4) the uncertain warm and cold dust emission in the EBL, which can be probed only by multi-TeV photons.
Hopefully the future TeV observatories will achieve substantially larger collection areas
as to dramatically improve the sensitivity and the quality of the spectra 
in this crucial energy range.

\section*{Acknowledgments}
I am grateful to G. Tosti and M. Ajello for helpful 
comments on the manuscript, and to F. Aharonian and D. Horns 
for years of fruitful discussions.

\end{document}